\documentclass[twocolumn]{aastex631}

\usepackage{amsmath}
\usepackage{xspace}

\newcommand{\um}{$\mu$m}

\newcommand{\htwo}{\ensuremath{\mathrm{H_2}}}
\newcommand{\hone}{\ion{H}{1}}
\newcommand{\hplus}{\ion{H}{2}}

\newcommand{\qpah}{\ensuremath{q_\mathrm{PAH}}}

\newcommand{\aco}{\ensuremath{\alpha_\mathrm{CO}}}

\newcommand{\ipah}{\ensuremath{I_\nu^\mathrm{F770W}}}
\newcommand{\idust}{\ensuremath{I_\nu^\mathrm{F2100W}}}
\newcommand{\iratio}{{\ensuremath{\ipah/\idust}}}

\usepackage{CJKutf8}

\begin{document}

\title{Relationships between PAHs, Small Dust Grains, \htwo, and \hone\ in Local Group Dwarf Galaxies NGC~6822 and WLM Using \textit{JWST}, ALMA, and the VLA}

\correspondingauthor{Ryan Chown}
\email{rchown53@gmail.com}

\newcommand{\Ox}{Sub-department of Astrophysics, Department of Physics, University of Oxford, Keble Road, Oxford OX1 3RH, UK}

\newcommand{\UGent}{Sterrenkundig Observatorium, Universiteit Gent, Krijgslaan 281 S9, B-9000 Gent, Belgium}

\newcommand{\STScI}{Space Telescope Science Institute, 3700 San Martin Drive, Baltimore, MD 21218, USA}

\newcommand{\MPIA}{Max-Planck-Institut f\"{u}r Astronomie, K\"{o}nigstuhl 17, D-69117, Heidelberg, Germany}

\newcommand{\AURA}{AURA for the European Space Agency (ESA), Space Telescope Science Institute, 3700 San Martin Drive, Baltimore, MD 21218, USA}

\newcommand{\UCSD}{Department of Astronomy \& Astrophysics, University of California, San Diego, 9500 Gilman Dr., La Jolla, CA 92093, USA}

\newcommand{\Whitman}{Whitman College, 345 Boyer Avenue, Walla Walla, WA 99362, USA}

\newcommand{\JHU}{Department of Physics and Astronomy, The Johns Hopkins University, Baltimore, MD 21218, USA}

\newcommand{\OSU}{Department of Astronomy, The Ohio State University, 140 West 18th Avenue, Columbus, OH 43210, USA}

\newcommand{\OSUPhys}{Department of Physics, The Ohio State University, Columbus, Ohio 43210, USA}

\newcommand{\CCAPP}{Center for Cosmology and Astroparticle Physics (CCAPP), 191 West Woodruff Avenue, Columbus, OH 43210, USA}

\newcommand{\ARI}{Astronomisches Rechen-Institut, Zentrum f\"{u}r Astronomie der Universit\"{a}t Heidelberg, M\"{o}nchhofstr. 12-14, D-69120 Heidelberg, Germany}

\newcommand{\ANU}{Research School of Astronomy and Astrophysics, Australian National University, Canberra, ACT 2611, Australia}

\newcommand{\UConn}{Department of Physics, University of Connecticut, 196A Auditorium Road, Storrs, CT 06269, USA}

\newcommand{\UHawaii}{Institute for Astronomy, University of Hawaii, 2680 Woodlawn Drive, Honolulu, HI 96822, USA}

\newcommand{\UniCA}{Universit\'{e} C\^{o}te d'Azur, Observatoire de la C\^{o}te d'Azur, CNRS, Laboratoire Lagrange, 06000, Nice, France}

\newcommand{\UAlberta}{Dept. of Physics, University of Alberta, 4-183 CCIS, Edmonton, Alberta, T6G 2E1, Canada}

\newcommand{\Arcetri}{INAF — Osservatorio Astrofisico di Arcetri, Largo E. Fermi 5, I-50125, Florence, Italy}

\newcommand{\UWyoming}{Department of Physics and Astronomy, University of Wyoming, Laramie, WY 82071, USA}

\newcommand{\LJMU}{Astrophysics Research Institute, Liverpool John Moores University, 146 Brownlow Hill, Liverpool L3 5RF, UK}

\newcommand{\ITA}{Universit\"{a}t Heidelberg, Zentrum f\"{u}r Astronomie, Institut f\"{u}r Theoretische Astrophysik, Albert-Ueberle-Str 2, D-69120 Heidelberg, Germany}

\newcommand{\CfA}{Center for Astrophysics $\mid$ Harvard \& Smithsonian, 60 Garden St., 02138 Cambridge, MA, USA}

\newcommand{\MPE}{Max-Planck-Institut f\"{u}r Extraterrestrische Physik (MPE), Giessenbachstr. 1, D-85748 Garching, Germany}

\newcommand{\UMD}{Department of Astronomy and Joint Space-Science Institute, University of Maryland, College Park, MD 20742, USA}

\newcommand{\Wisc}{University of Wisconsin–Madison, Department of Astronomy, 475 N Charter St, Madison, WI 53703, USA}

\newcommand{\UVA}{Department of Astronomy, University of Virginia, Charlottesville, VA, USA}

\newcommand{\NRAO}{National Radio Astronomy Observatory, Charlottesville, VA, USA}

\newcommand{\ASIAA}{Institute of Astronomy and Astrophysics, Academia Sinica, No. 1, Sec. 4, Roosevelt Road, Taipei 106216, Taiwan}

\newcommand{\kipac}{Kavli Institute for Particle Astrophysics \& Cosmology (KIPAC), Stanford University, CA 94305, USA}

\newcommand{\UBonn}{Argelander-Institut f\"ur Astronomie, Universit\"at Bonn, Auf dem H\"ugel 71, 53121 Bonn, Germany}

\newcommand{\UWyo}{Department of Physics and Astronomy, University of Wyoming, Laramie, WY 82071, USA}

\newcommand{\uconcepcion}{Departamento de Astronom\'ia, Universidad de Concepci\'on, Barrio Universitario, Concepci\'on, Chile}

\newcommand{\UW}{Department of Astronomy, University of Washington, Seattle, WA, 98195}

\newcommand{\cca}{Center for Computational Astrophysics, Flatiron Institute, 162 Fifth Ave, New York, NY, 10010, USA}

\author[0000-0001-8241-7704]{Ryan~Chown}
\affiliation{\OSU}

\author[0000-0002-2545-1700]{Adam~K.~Leroy}
\affiliation{\OSU}
\affiliation{\CCAPP}

\author[0000-0002-5480-5686]{Alberto~D.~Bolatto}
\affiliation{\UMD}

\author[0000-0002-5235-5589]{J\'{e}r\'{e}my~Chastenet}
\affiliation{\UGent}

\author[0000-0001-6708-1317]{Simon~C.~O.~Glover}
\affiliation{\ITA}

\author[0000-0002-4663-6827]{R\'{e}my~Indebetouw}
\affiliation{\UVA}
\affiliation{\NRAO}

\author[0000-0001-9605-780X]{Eric~W.~Koch}
\affiliation{\CfA}

\author[0000-0002-3106-7676]{Jennifer Donovan Meyer}
\affiliation{\NRAO}

\author[0000-0001-9504-7386]{Nickolas M. Pingel}
\affiliation{\Wisc}

\author[0000-0002-5204-2259]{Erik~Rosolowsky}
\affiliation{\UAlberta}

\author[0000-0002-4378-8534]{Karin~Sandstrom}
\affiliation{\UCSD}

\author[0000-0002-9183-8102]{Jessica~Sutter}
\affiliation{\Whitman}

\author[0000-0003-1356-1096]{Elizabeth Tarantino}
\affiliation{\STScI}

\author[0000-0003-0166-9745]{Frank Bigiel}
\affiliation{\UBonn}

\author[0000-0003-0946-6176]{Médéric Boquien}
\affiliation{\UniCA}

\author[0000-0003-2551-7148]{I-Da Chiang \begin{CJK*}{UTF8}{bkai}(江宜達)\end{CJK*}}
\affiliation{\ASIAA}

\author[0000-0002-5782-9093]{Daniel~A.~Dale}
\affiliation{\UWyo}

\author[0000-0002-1264-2006]{Julianne J.\ Dalcanton}
\affiliation{\cca}
\affiliation{\UW}

\author[0000-0002-4755-118X]{Oleg V. Egorov}
\affiliation{\ARI}

\author[0000-0002-1185-2810]{Cosima Eibensteiner}
\altaffiliation{Jansky Fellow of the National Radio Astronomy Observatory}
\affiliation{\NRAO}

\author[0000-0002-3247-5321]{Kathryn~Grasha}
\affiliation{\ANU}

\author[0000-0002-8806-6308]{Hamid~Hassani}
\affiliation{\UAlberta}

\author[0000-0001-9020-1858]{Hao He}
\affiliation{\UBonn}

\author[0000-0002-0432-6847]{Jaeyeon Kim}
\affiliation{\kipac}

\author[0000-0002-6118-4048]{Sharon~Meidt}
\affiliation{\UGent}

\author[0000-0003-2721-487X]{Debosmita Pathak}
\affiliation{\OSU}
\affiliation{\CCAPP}

\author[0000-0002-4781-7291]{Sumit K. Sarbadhicary}
\affiliation{\OSUPhys}
\affiliation{\CCAPP}
\affiliation{\OSU}

\author[0000-0002-3418-7817]{Snezana Stanimirovic}
\affiliation{\Wisc}

\author[0000-0002-5877-379X]{Vicente Villanueva}
\affiliation{\uconcepcion}

\author[0000-0002-0012-2142]{Thomas G. Williams}
\affiliation{\Ox}

\begin{abstract}
We present 0.7--3.3 pc resolution mid-infrared (MIR) JWST images at 7.7~\um\ (F770W) and 21~\um\ (F2100W) covering the main star-forming regions of two of the closest star-forming low-metallicity dwarf galaxies, NGC~6822 and Wolf–Lundmark–Melotte (WLM). The images of NGC~6822 reveal filaments, edge-brightened bubbles, diffuse emission, and a plethora of point sources. By contrast, most of the MIR emission in WLM is point-like, with a small amount of extended emission. Compared to solar metallicity galaxies, the ratio of 7.7~\um\ intensity (\ipah), tracing polycyclic aromatic hydrocarbons (PAHs), to 21~\um\ intensity (\idust), tracing small, warm dust grain emission, is suppressed in these low-metallicity dwarfs. Using ALMA CO(2-1) observations, we find that detected CO intensity versus \ipah\ at $\approx 2$~pc resolution in dwarfs follows a similar relationship to that at solar metallicity and lower resolution, while the CO versus \idust\ relationship in dwarfs lies significantly below that derived from solar metallicity galaxies at lower resolution, suggesting more pronounced destruction of CO molecules at low metallicity. Finally, adding in Local Group L-Band Survey VLA 21-cm \hone\ observations, we find that \idust\ and \ipah\ vs. total gas ratios are suppressed in NGC~6822 and WLM compared to solar metallicity galaxies. In agreement with dust models, the level of suppression appears to be at least partly accounted for by the reduced galaxy-averaged dust-to-gas and PAH-to-dust mass ratios in the dwarfs. Remaining differences are likely due to spatial variations in dust model parameters, which should be an exciting direction for future work in local dwarf galaxies.
\end{abstract}


\section{Introduction} \label{sec:intro}

Low-metallicity star-forming regions (hereafter ``SF regions'') are common in outer disks of galaxies like the Milky Way, in dwarf galaxies, and at high redshifts \citep{maiolino2019}. The low metallicity in these regions influences the star formation process and interstellar medium (ISM) conditions. Metallicity ($Z$) governs the abundances of dust and ISM coolants -- including CO, the most widely-used molecular gas tracer \citep{bolatto2013}. This, in turn, affects the balance between atomic and molecular gas \citep{glover2012, galliano2018, saintonge2022}. 

The nearest accessible extragalactic low-$Z$ SF regions are found in dwarf galaxies\footnote{Throughout this work, by ``dwarf galaxies'' or ``dwarfs'' we mean ``dwarf irregulars'' as opposed to dwarf galaxies with little-to-no interstellar medium, e.g., dwarf spheroidals.}, which have low total masses, high \hone\ mass fractions, low dust-to-gas mass ratios (DGRs), harder radiation fields, and lower molecular-to-total hydrogen ratios compared to the metal-rich (e.g., solar metallicity) massive star-forming galaxies \citep[for a review see][]{hunter2024}. 

While we have theoretical predictions of the structure and physical state of the gas in low-$Z$ SF regions \citep{glover2012b, hu2021, hu2023, kim2024}, observations that test these predictions have been challenging \citep[see summary in][]{hunter2024}. To a large extent, this reflects that measuring the amount of \htwo\ in low-$Z$ SF regions is challenging. Millimeter-wavelength rotational line emission from carbon monoxide (CO), the most widely-used tracer of \htwo, becomes faint at low metallicity \citep{maloney1988,grenier2005,wolfire2010,glover2012,schruba2012, bolatto2013, madden2020}. At solar metallicity ($Z=Z_\odot$), dust shields CO from dissociating far ultraviolet radiation, whereas at $Z\leq 0.3~Z_\odot$, CO is preferentially dissociated relative to \htwo, which self-shields much more effectively \citep{madden2020}. As a result, low metallicity molecular clouds are expected to harbor significant amounts \citep[$\gtrsim 70$~\%;][]{madden2020} of \htwo\ that is not traceable by CO \citep[i.e., it is ``CO-dark'';][]{maloney1988,grenier2005,wolfire2010,glover2012}. At very low metallicities (about $0.1~Z_\odot$ or lower) up to 100\% of the \htwo\ mass may be CO-dark \citep[e.g.,][]{shi2016, madden2020, shi2020}. 

Attempting to measure the amount and properties of CO-dark \htwo\ is an active area of research. Spectroscopy of multiple carbon species \citep[e.g.,][]{pineda2017,jameson2018}, dynamical measurements based on the detected CO, UV absorption or IR emission tracing H$_2$ directly, and even gamma rays all offer possible routes forward \citep[see review in][]{bolatto2013}. Perhaps the most widely-used approach is to use observations of dust to trace the gas \citep[e.g.,][]{israel1997,leroy2011}. This approach assumes that dust and gas are well-mixed, and leverages the fact that dust emission is often still visible in emission even at low $Z$. 

Spatial resolution has been a major obstacle to using dust to trace \htwo\ in dwarf galaxies. Far-IR telescopes have had limited angular resolution (e.g., \textit{Herschel} had FWHM~$\approx 13''$ near the peak of the IR SED at 160~$\mu$m) and the molecular clouds in dwarf galaxies tend to be small \citep[e.g.,][]{rubio2015,schruba2017,shi2020}. JWST can map mid-IR (MIR) emission at much higher resolution, $\sim 0.3{-}1''$, offering the prospect to push dust studies of low-$Z$ SF regions into a new era, even resolving clouds in the Local Group where $1'' \lesssim 5$~pc. 

To help take this next step, we present new JWST MIRI imaging of MIR dust emission from five low-$Z$ SF regions in two Local Group galaxies, NGC~6822 and the Wolf-Lundmark-Melotte (WLM) galaxy (Table~\ref{tab:gal}). Along with NGC~2366 \citep{oey2017} and Sextans~B \citep{shi2020}, NGC~6822 \citep{schruba2017} and WLM \citep{rubio2015} are the only low-$Z$ galaxies beyond the Small Magellanic Cloud (SMC) where CO emission has been detected and resolved into individual clouds and cores \citep{hunter2024}. Therefore, these regions harbor reservoirs of molecular gas in addition to signatures of recent high mass star formation. 

An important difference from previous efforts using dust to trace CO-dark H$_2$ is that JWST observes the MIR. The MIR is dominated by continuum emission from stochastically-heated small dust grains and a number of strong, broad emission features produced by the stretching and bending modes of polycyclic aromatic hydrocarbons \citep[PAHs; ][]{allamandola1989, tielens2008, draine2007a, draine2011, galliano2018, li2020, hensley2022, hensley2023}. Neither MIR continuum nor PAH emission directly trace the total dust mass in the same way as the optical depth derived from the SED \citep[e.g.,][]{galliano2018}, but PAH and stochastic emission from small dust grains are expected to trace the product of gas column density, the intensity of the interstellar radiation field, and the small grain abundance. In more massive galaxies, there are strong correlations between this MIR dust emission and CO emission on scales as large as whole galaxies and small as molecular clouds \citep[e.g.][]{gao2019, chown2021, leroy2021, leroy2023, sandstrom2023a, whitcomb2023, chown2025, villanueva2025}. 

This work aims to explore how the use of MIR emission to trace gas extends to low-$Z$ SF regions. Except for a pioneering study by \citet{gratier2010}, this relationship is mostly unexplored in low-$Z$ galaxies. One clear issue is variations in PAH abundance. Compared to massive, metal-rich galaxies, low-$Z$ galaxies exhibit low PAH abundances as traced by both SED fitting and the ratio of PAH-to-small dust grain emission \citep{engelbracht2005, hogg2005, madden2006, gordon2008, draine2007,li2020a, chastenet2025}. This reflects some combination of enhanced PAH destruction and reduced PAH production at low $Z$. Resolved mapping of PAH abundances in the Magellanic Clouds suggested an overall drop in PAH abundance near the metallicity of the SMC, enhanced PAH abundance in molecular gas-dominated regions, and suppressed PAH abundance in \ion{H}{2} regions \citep{sandstrom2010,chastenet2019}. JWST allows similar resolved measurements beyond the Magellanic Clouds for the first time. With these goals in mind we address the following questions.

\begin{figure*}[t!]
\begin{center}
\includegraphics[width=\textwidth]{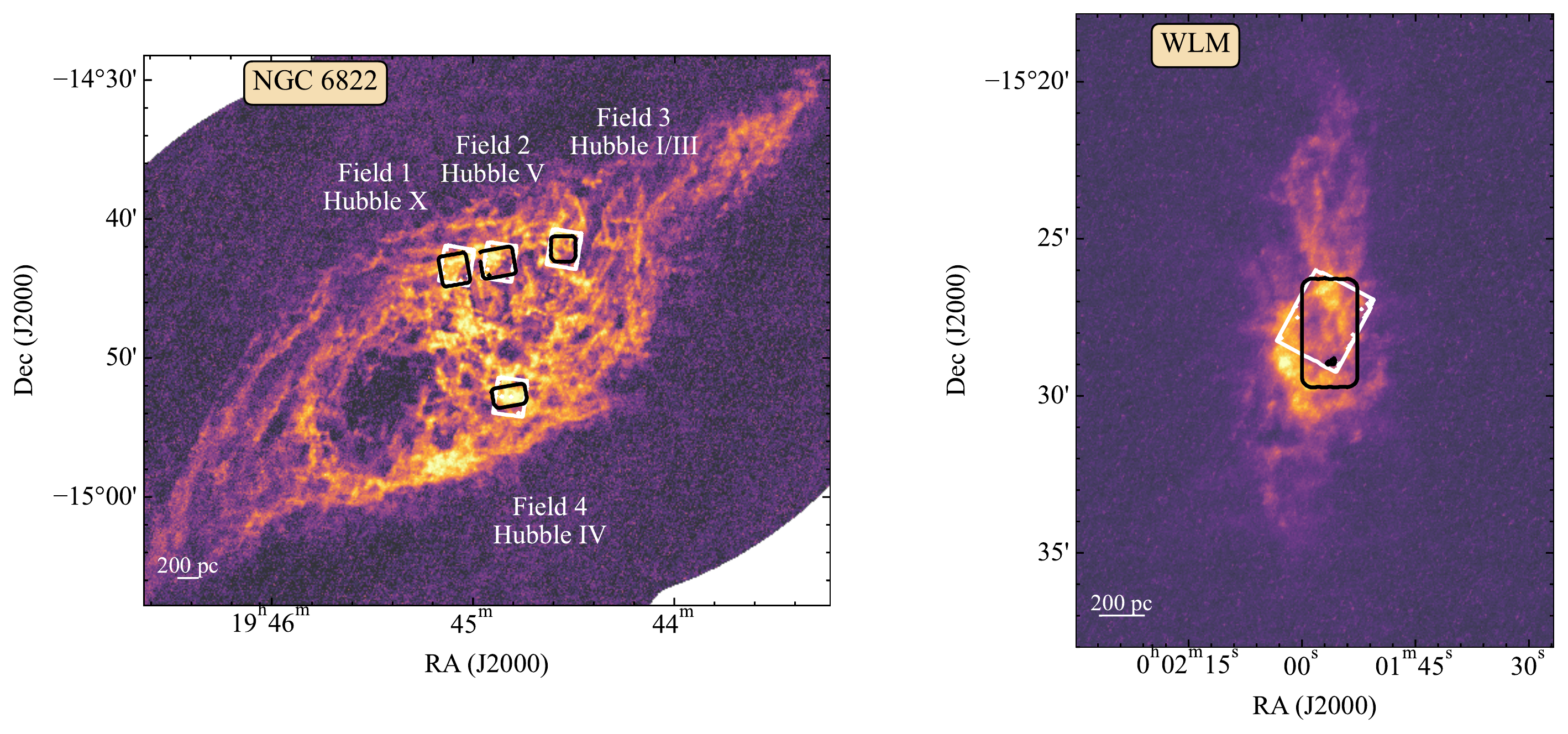}
\caption{$4\arcsec$ 21~cm~\hone\ peak intensity image from the Local Group L-Band Survey (E. Koch et al. submitted, \citealt{pingel2024}, N. Pingel et al., in preparation). Contours show the footprints of our JWST MIRI observations (white) and ALMA CO(2-1) data (black). 
\label{fig:overview}}
\end{center}
\end{figure*}

\begin{enumerate}
\item How does the ratio of PAH emission to small dust grain emission -- a proxy for the PAH-to-dust mass fraction -- vary across these regions? In particular, how does this ratio vary between \ion{H}{2} regions and the surrounding molecular clouds? How do the ratios compare between the low metallicity NGC~6822, the very low metallicity WLM, and more massive, metal rich galaxies?

\item In the regions that do show CO emission, how does the CO emission observed by ALMA correlate with the PAH and small dust grain emission ratios? How do the normalization and strength of these correlations compare to those found in massive, high metallicity galaxies?

\item How do PAH and small dust grain emission correlate with ISM tracers, especially the atomic gas traced by 21-cm mapping, over the full field of our images? Is there evidence for a phase dependence of the small grain or PAH abundance?
\end{enumerate}

We describe our new observations and supporting data in \S \ref{sec:data}, address these questions \S \ref{sec:results} and summarize our conclusions in \S \ref{sec:conclusions}. In Appendix~\ref{sec:appendix_qpah} we connect our MIR band ratios to dust model parameters and discuss implications. 

\section{Data}
\label{sec:data}

\begin{deluxetable*}{lllll}
\tablecolumns{5}
 \tablecaption{Target galaxy properties.
    \tablehead{
         \colhead{Property} & \colhead{NGC~6822} & \colhead{Reference} & \colhead{WLM} & \colhead{Reference}}}
\startdata
Hubble type & IB(s)m & NED & IB(s)m & NED \\
R.A. (J2000) & 19h44m55.74s & NED & 00h01m58.1610s & NED \\
Dec. (J2000) & $-$14d48m12.4s & NED & $-$15d27m39.340s & NED \\
Distance (kpc) & $474\pm 13$ & \citet{rich2014} & $985 \pm 33$ & \citet{leaman2012} \\
$Z/Z_\odot$\tablenotemark{a}& 0.2 & \citet{hernandez-martinez2009} & 0.13 & \citet{lee2005}\\
$M_*/M_\odot$ & $1.5\times 10^8$ & \citet{madden2014} & $1.6  \times 10^{7}$ & \citet{zhang2012} \\
$M_\mathrm{HI}/M_\odot$ & $1.3\times 10^8$ & \citet{weldrake2003} &  $3.2\pm 0.3\times 10^7$ & \citet{jackson2004}  \\
DGR\tablenotemark{b} & $\approx 2.0 \times 10^{-3}$ & \citet{schruba2017} & $\approx 1\times 10^{-3}$ & \citet{remy-ruyer2014} \\
SFR/M$_\odot$~yr$^{-1}$ & $0.015$  & \citet{efremova2011} & $0.006$ & \citet{hunter2010} \\ 
$\alpha_\mathrm{CO}/\alpha_\mathrm{CO,MW}$\tablenotemark{b} & 31 & Equation~\ref{eq:aco} & 60 & Equation~\ref{eq:aco} \\
\cutinhead{RMS noise in F770W images at varying resolutions [MJy~sr$^{-1}$]}
Native & 0.062 & & 0.041 & \\
$0.9''$ & 0.022 & & 0.010 & \\
$2.0''$ & 0.019 & & 0.007 & \\
\cutinhead{RMS noise in F2100W images at varying resolutions [MJy~sr$^{-1}$]}
Native & 0.283 & & 0.233 & \\
$0.9''$ & 0.085 & & 0.058 & \\
$2.0''$ & 0.059 & & 0.025 & \\
\enddata
\tablecomments{\tablenotetext{a}{Total gas-phase oxygen abundance in \hplus\ regions, converted to $Z$ assuming $12+\log\mathrm{O/H} = \log_{10}(Z/Z_\odot) + 8.73~\mathrm{dex}$. Note that we expect region-to-region variations in $Z$ \citep[e.g., as in the SMC/LMC,][]{choudhury2016, choudhury2018} but lack measurements for each region.
}
\tablenotetext{b}{For NGC~6822 we show 1/GDR from Table 1 of \citet{schruba2017}, where the dust mass is from \citet{remy-ruyer2015} and the gas mass is from \citet{weldrake2003} and \citet{gratier2010}. For WLM we use the broken powerlaw prescription for GDR assuming $Z=0.13~Z_\odot$ and a metallicity dependent CO-to-H$_2$ conversion factor \citep[Table 1][]{remy-ruyer2014}. We also use this broken powerlaw to estimate DGR for PHANGS (\S\ref{sec:mir_gas}).}
\tablenotetext{c}{CO-to-H$_2$ conversion factor relative to the Milky Way value $\alpha_\mathrm{CO,MW}=4.35$~M$_\odot$~pc$^{-2}$~(K~km~s$^{-1}$)$^{-1}$, computed via Equation~\ref{eq:aco}.}
}
\label{tab:gal}
\end{deluxetable*}

\subsection{New \textit{JWST} 7.7 and 21 $\mu$m observations} 
\label{subsec:data_jwst}

As part of Cycle 2 GO program \# 4256, we used JWST MIRI to image four star-forming complexes in NGC~6822 and one in WLM (Figure~\ref{fig:overview}). CO(2-1) emission has been resolved and detected in each of these regions using ALMA (\S\ref{subsec:data_alma}). The complexes collectively capture most of the star formation activity in both galaxies. Fields 1 (Hubble X), 2 (Hubble V), 3 (Hubble I/III), and 4 (Hubble IV) collectively account for $\approx 2/3$ of the global H$\alpha$ and Spitzer 24~\um\ fluxes in NGC~6822 \citep{schruba2017}. 

The observations occurred on 19 October and 01 November 2023. We observed each field using the F770W and F2100W filters and a $2 \times 1$ pointing mosaic, which covered the CO and mid-IR emission known from previous observations \citep[][]{cannon2006}, and most of the associated H$\alpha$ emission \citep[][]{hunter2012,schruba2017}.  We used the FASTR1 readout pattern, with 20 groups per integration and a 4-point dither. The F770W observations were done with 1 integration per exposure, while F2100W were done with 2 integrations per exposure. We observed one off-galaxy field each for NGC~6822 and WLM, and used this to construct a background that we subtracted from each on-source field. The total exposure times for each $2\times1$ field mosaic were 444~s for F770W and 910~s for F2100W. The corresponding per-point exposure times were therefore 222~s for F770W and 455~s for F2100W.  

The resolution of JWST is $0.269\arcsec$ at F770W and $0.674\arcsec$ at F2100W. At the distance of NGC~6822 (Table~\ref{tab:gal}), the corresponding linear resolutions are 0.62~pc (F770W) and 1.55~pc (F2100W), while for WLM this corresponds to linear resolutions of 1.28~pc (F770W) and 3.22~pc (F2100W). The $\approx 200 \arcsec \times 125\arcsec$ extent of our mosaics corresponds to an areal coverage of $\approx 500~\mathrm{pc} \times 300~\mathrm{pc}$ for NGC~6822, and $\approx 950~\mathrm{pc} \times 600~\mathrm{pc}$ for WLM.

The data were reduced using \texttt{pjpipe} \citep{williams2024} version 1.2.0, JWST pipeline (\texttt{jwst}) version 1.17.1, and CRDS context \texttt{1322.pmap}. Compared to the current default observatory pipeline, \texttt{pjpipe} implements improved solutions for matching backgrounds between different tiles of the mosaics and improves the astrometric alignment between dithers and adjacent tiles. We find that individual exposures in dither sequences show varying background levels. The origins of these offsets may be due to persistence \citep{morrison2023, dicken2024}, but at the time of writing this is not completely certain. These offsets present a challenge for this data set in particular because the targeted regions show faint extended emission with surface brightness $I_\nu \approx  0.1$~MJy~sr$^{-1}$, while the individual exposures can differ by a similar amount in overlapping areas. 
This problem is more significant for the F2100W observations, where the backgrounds are much brighter than for F770W \citep{rigby2023}. We use the \texttt{skymatch} step of the JWST pipeline with \texttt{method=local}, which helps to make the overall flux level of exposures and adjacent pointings more uniform. We estimate the uncertainty on the overall intensity level to be about $0.1$~MJy~sr$^{-1}$.

In order to compare the JWST bands to one another and to other data with different PSFs, we convolve the maps to several  resolutions. We construct kernels that convert from JWST PSFs generated using \texttt{webbpsf}\footnote{\url{https://webbpsf.readthedocs.io; XXX}} to a series of Gaussian PSFs with different full widths at half-maximum (FWHMs). We construct versions of the JWST maps with FWHM $0.9\arcsec$ ($\approx 2.1$~pc at NGC~6822 and $4.3$~pc at WLM). As discussed in \citet{williams2024}, this represents the sharpest Gaussian PSF to which the F2100W data can be safely convolved and this convolution decreases the noise of F2100W images by a factor of about $\approx 3$. We also create versions of the maps at the $2\arcsec$ ($\approx 4.7$~pc or $9.6$~pc) working resolutions of the ALMA CO data (see Section~\ref{subsec:data_alma}) and the $7.5\arcsec$ (NGC~6822) and $8.25\arcsec$ (WLM) resolution of the VLA 21-cm maps (Section~\ref{subsec:data_vla}).

The noise level in the maps depends on resolution. Table~\ref{tab:gal} reports the statistical noise in each filter at several resolutions, gauged from apparently signal-free regions of the images (the rms noise reported for NGC~6822 is the average of the rms noise levels in Fields 1--4)

\begin{figure*}
\begin{center}
\includegraphics[width=\textwidth]{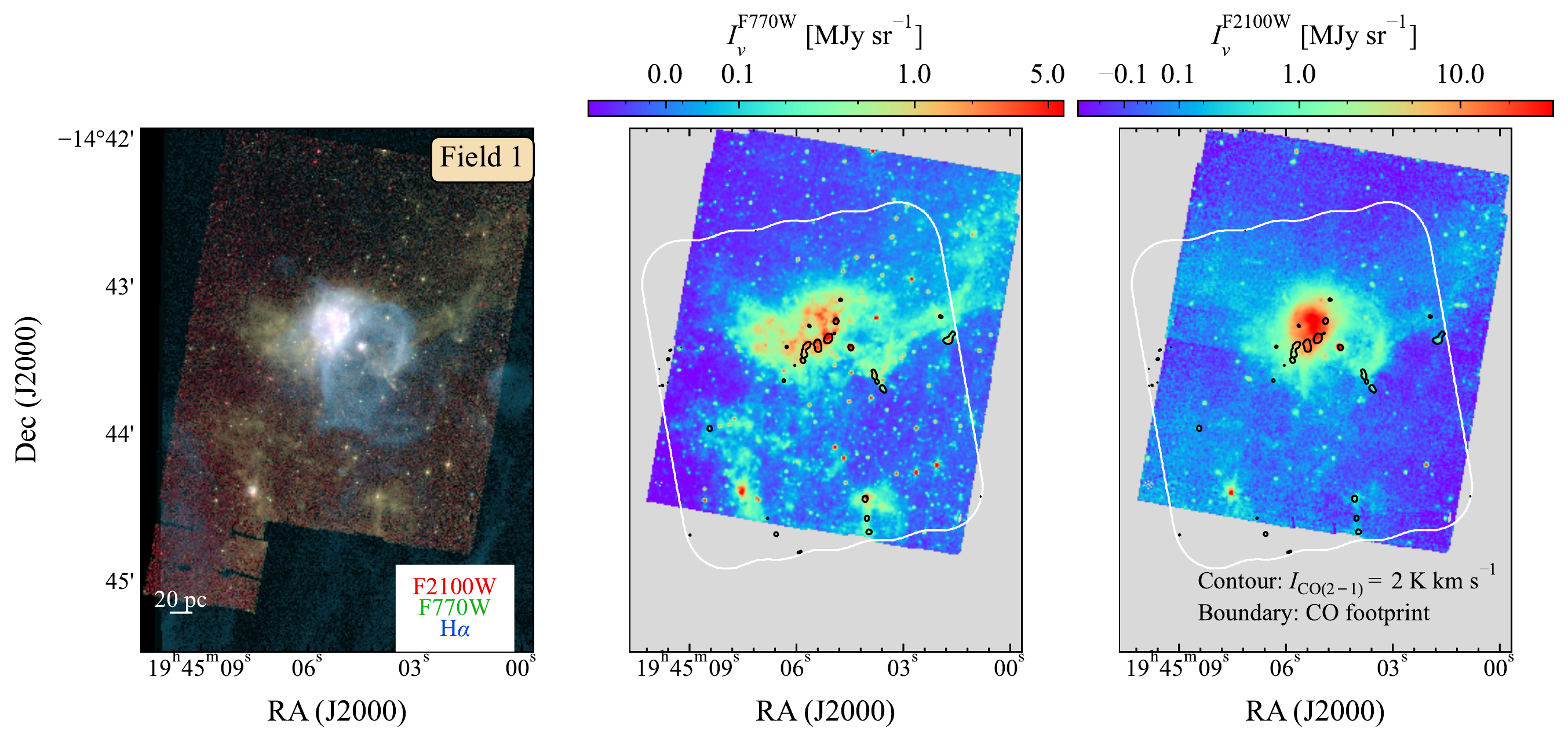}
\includegraphics[width=\textwidth]{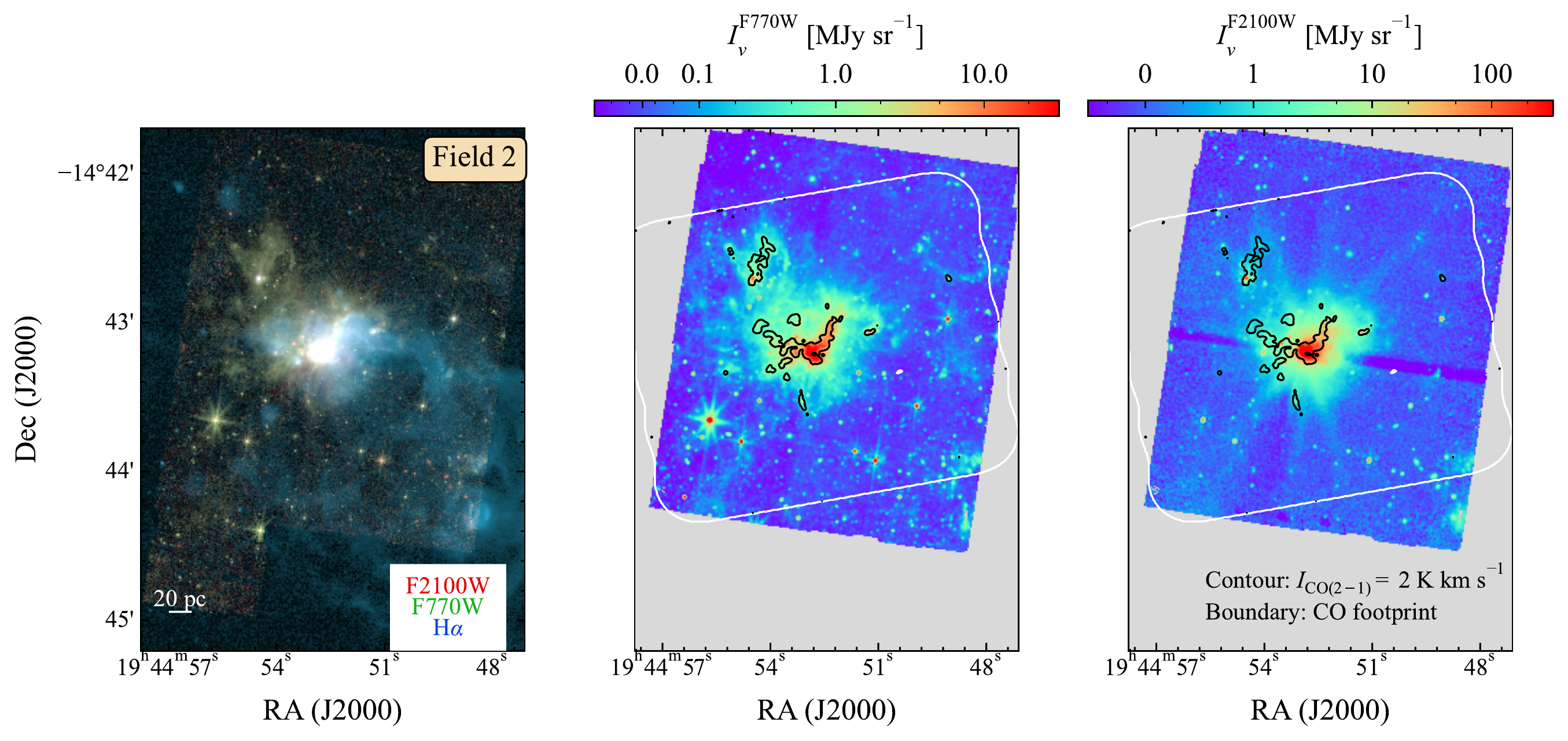}
\caption{Left: images of Fields 1 (top) and 2 (bottom) in NGC~6822 combining JWST F2100W (red), JWST F770W (green), and H$\alpha$ (blue) from SIGNALS. Middle and right: the F770W and F2100W maps from JWST at $0.9''$ resolution. The white boundary shows the field of view of ALMA CO(2-1) mapping (Sec.~\ref{subsec:data_alma}), and the black contours show where CO is well-detected. The diffraction spikes are artifacts around point sources in both F770W and F2100W. Note how MIR emission is detected over a much more extended area than CO. \label{fig:rgb_6822_1}}
\end{center}
\end{figure*}

\begin{figure*}
\begin{center}
\includegraphics[width=\textwidth]{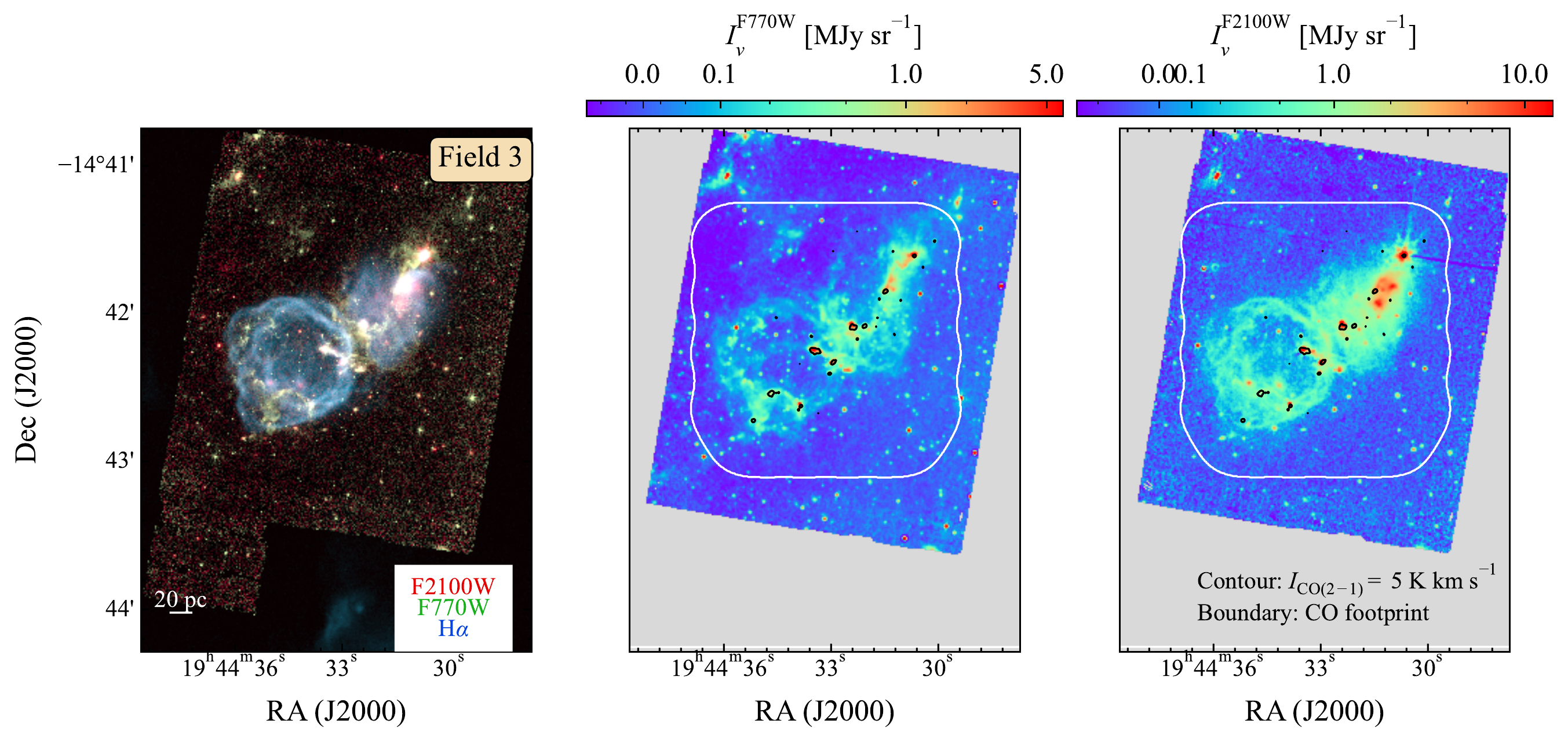}
\includegraphics[width=\textwidth]{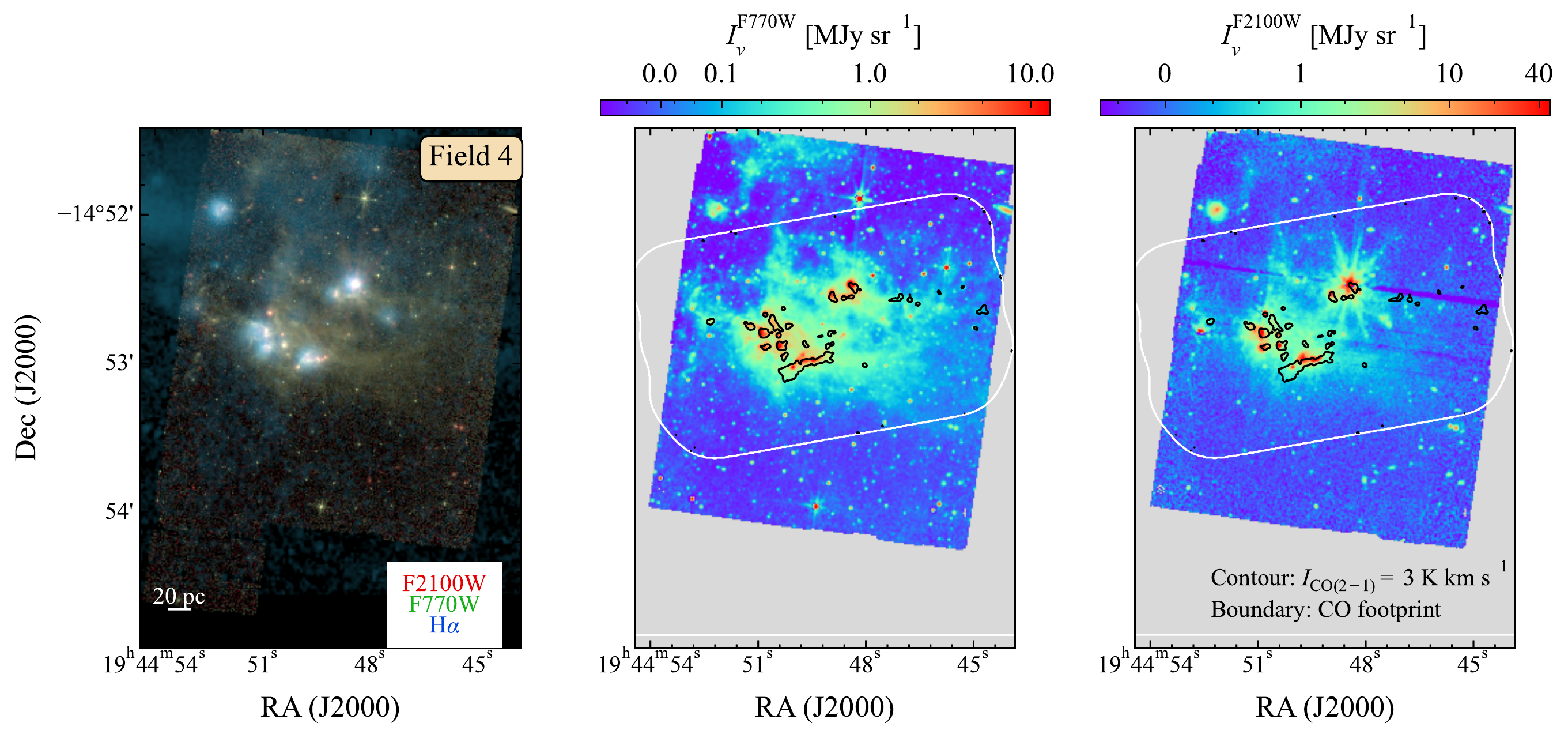}
\caption{As Fig. \ref{fig:rgb_6822_1} but for NGC 6822 Fields 3 (top) and 4 (bottom). In Field 4 the H$\alpha$ data come from \citet{hunter2012}. \label{fig:rgb_6822_3}}
\end{center}
\end{figure*}

\begin{figure*}
\begin{center}
\includegraphics[width=\textwidth]{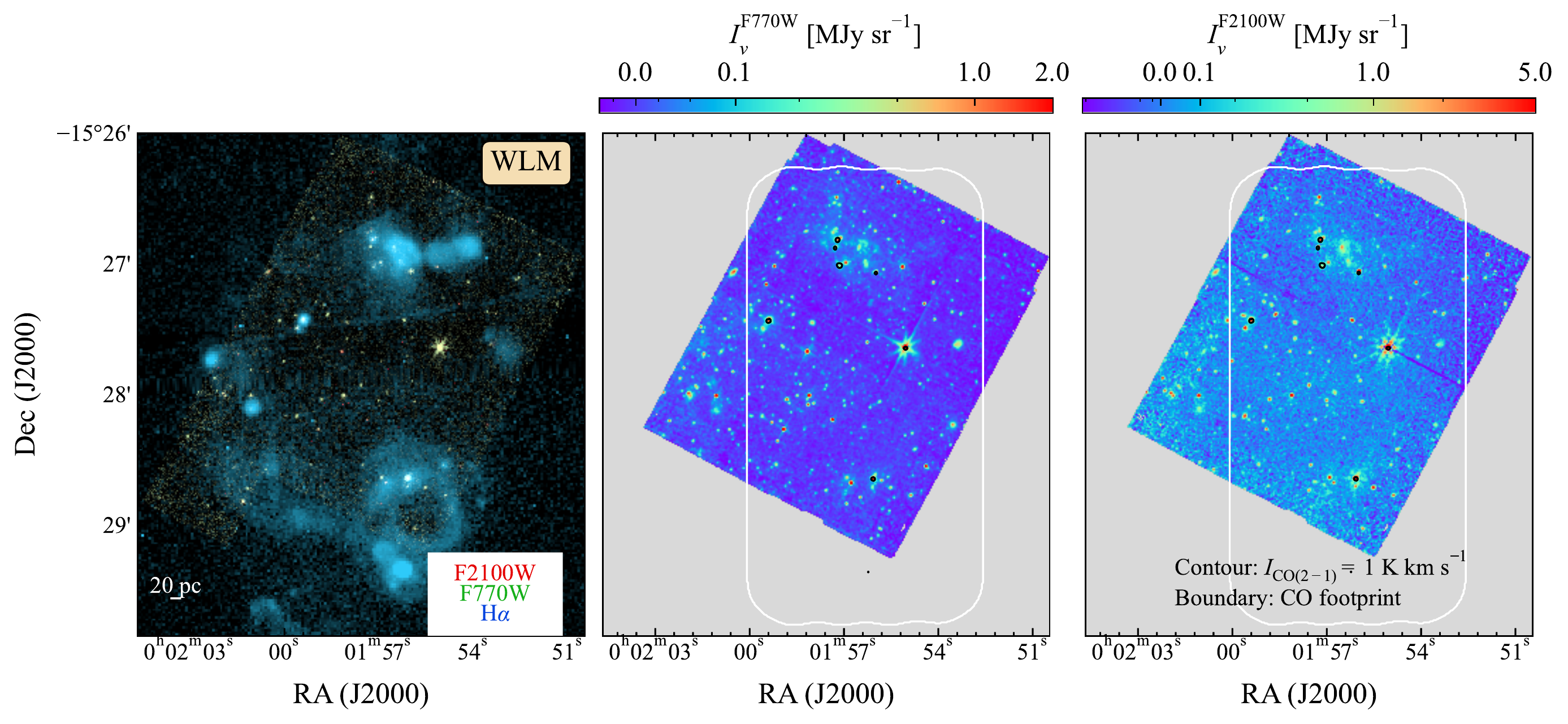}
\caption{As Fig. \ref{fig:rgb_6822_1} but for WLM. The H$\alpha$ here comes from \citet{hunter2012}.
\label{fig:rgb_wlm}}
\end{center}
\end{figure*}

\subsection{ALMA CO(2-1)} \label{subsec:data_alma}

We compare MIR emission to ALMA CO(2-1) data. For WLM these are from ALMA project 2018.1.00337.S (PI Rubio). They have appeared in \citet{archer2024} and will be presented in M. Rubio et al.\ (in preparation). For NGC~6822 we use the CO~(2-1) data from \citet[][]{schruba2017} for Field 3. For Fields 1, 2, and 4 we use new CO~(2-1) mapping from project 2024.1.01179.S (PI Chown). These new data cover approximately the same area as the \citet{schruba2017} mapping of these fields but achieve $\approx 3{-}4\times$ better sensitivity. This project will eventually include short and zero spacing data, but these are still being observed. Here we use the observatory-delivered main array (12-m) imaging only. In both galaxies, the CO mosaics approximately match the coverage of our JWST data. We convolve the data to have round synthesized beams of $1.35\arcsec$ for NGC~6822 Field 1, $1.2''$ for Field 2, $1.1''$ for Fields 3 and Field 4; and $2\arcsec$ for WLM. The maximum recoverable angular scale is $11''$ for NGC~6822 and $16''$ for WLM.

We produce integrated intensity maps at the native, $2\arcsec$ (FWHM) resolution, and resolution matched to the 21-cm data. To do this, we integrated over a velocity range of $\pm 8.4$~km~s$^{-1}$ around the local mean 21-cm velocity. The 21-cm line is detected everywhere in our fields and its velocity typically aligns well with the velocity of CO emission \citep[e.g.,][]{schruba2011}. We arrived at the $\pm 8.4$~km~s$^{-1}$ window by inspecting the CO cubes by eye and determining that the observed bright CO emission lies within this velocity range. After convolution to a matched $2''$ resolution, the resulting integrated intensity maps have typical uncertainties of $1\sigma \approx 0.1$~K~km~s$^{-1}$ for NGC~6822 Fields 1 and 2, $\approx 0.4$~K~km~s$^{-1}$ for NGC~6822 Field 3, and $\approx 0.15$~K~km~s$^{-1}$ for Field 4 and WLM. 

We analyze the integrated intensity of CO(2-1) rather than \htwo\ surface density except for Sec.~\ref{sec:mir_gas} where we combine CO and \hone\ measurements to calculate total H column densities. To do this, we employ CO-to-\htwo\ conversion factors, \aco , that assume a scaling with metallicity \citep[following][]{schinnerer2024}
\begin{equation}\label{eq:aco}
\aco=\alpha_\mathrm{CO,MW} \left(Z/Z_\odot \right)^{-1.5},
\end{equation}
using $Z$ in Table~\ref{tab:gal} \citep[also see][]{bisbas2025}. This yields $\aco$ of 134 and 261 $M_\odot$~pc$^{-2}(\mathrm{K~km~s^{-1}})^{-1}$ for NGC~6822 and WLM respectively. 

\subsection{VLA 21-cm} \label{subsec:data_vla}

Neutral atomic hydrogen (\ion{H}{1}) makes up most of the ISM in these galaxies. To trace this gas we use new VLA maps of the 21-cm line from the ``Local Group L Band Legacy Survey\footnote{\url{www.lglbs.org}}'' (LGLBS, E. Koch et al. ApJS submitted) and imaged as described in \citet[see][]{pingel2024}. We use 21-cm images that combined data from all four VLA configurations and the short-spacing data from the Green Bank Telescope. These images appear in Figure \ref{fig:overview}. We convolve both VLA data cubes to have a circular synthesized beam, which has FWHM $7.5\arcsec$ for NGC~6822 and $8.25\arcsec$ for WLM. We created line-integrated intensity masks for each field and converted these to column density assuming optically thin 21-cm emission\footnote{\citet{pingel2024} show that opacity can be significant within the star forming complexes in NGC~6822, but we lack a method to correct the \hone\ over each whole field. We expect the largest opacities to be present in the high column density regions near the centers of the complexes.}. The typical $1\sigma$ N(\ion{H}{1}) column density sensitivity of these maps is $< 5 \times 10^{19}$~cm$^{-2}$, and 21-cm emission is detected at good significance along all sight lines in our fields.

\subsection{H$\alpha$} 
\label{subsec:data_ha}

We use H$\alpha$ images to estimate the boundaries of the \hplus\ region in each field. Fields 1--3 in NGC~6822 were observed using the SITELLE optical IFU at the Canada-France-Hawaii Telescope (CFHT) as part of the SIGNALS survey \citep{rousseau-nepton2019}, and narrowband H$\alpha$ maps for all four NGC~6822 fields and WLM were obtained as part of the LITTLE THINGS survey \citep{hunter2012}. To be consistent across fields, we use the LITTLE THINGS narrowband images when analyzing H$\alpha$ intensity measurements, but use SIGNALS where available for the visualizations in Figures~\ref{fig:rgb_6822_1} -- \ref{fig:rgb_6822_3}. The map units for the LITTLE THINGS data were converted to flux from native units of counts by applying calibration factors available on the LITTLE THINGS website.\footnote{E.g.\ \url{http://www2.lowell.edu/users/dah/littlethings/wlm.html} for WLM.}
The seeing for the LITTLE THINGS H$\alpha$ data is $2.5\arcsec$ NGC~6822 and $3.2\arcsec$ for WLM \citep{hunter2004}. Further details of the LITTLE THINGS H$\alpha$ data can be found in \citet{hunter2004}.

\subsection{Comparison data}\label{subsec:compdata}

We compare our measurements for NGC~6822 and WLM to similar measurements for more massive galaxies at coarser resolution as well as to similar measurements of the Magellanic Clouds obtained with \textit{Spitzer}.

\subsubsection{Massive galaxies from PHANGS}\label{subsec:phangs}

We compare our measurements to F770W, F2100W, CO~(2-1), \ion{H}{1}, and H$\alpha$ observations of star-forming regions obtained by the PHANGS surveys \citep[see][]{leroy2021}. We use F770W and F2100W observations from PHANGS-JWST \citep[][Cycle 1 GO program 2107]{lee2023}. The processing for those data is described by \cite{williams2024}, and we use public release version v1.0.1. The CO~(2-1) observations come from PHANGS--ALMA \citep{leroy2021}. The extinction-corrected H$\alpha$ measurements are from PHANGS-MUSE \citep{emsellem2022}, and the identification of nebulae in PHANGS-MUSE are from \citet{groves2023}. The 21-cm data come from the compilations described in \citet{sun2020,sun2022,chiang2024}. 

Currently, the matched data described above exist for $19$ galaxies, which form our working comparison data set. Except for the 21-cm data, most data have angular resolution in the range $\approx 0.8{-}1.5\arcsec$, which corresponds to a common physical resolution $\approx 100{-}150$~pc at the distances to the galaxies. To form our comparison measurements, we convolve the ALMA, JWST, and MUSE data to the common angular resolution achievable for all three data sets for each galaxy. Then we project the maps to a common astrometric grid with Nyquist sampled pixels (i.e., pixel size $=\mathrm{FWHM}/2$) at this working resolution, and measure the intensities for each band at each location in each galaxy. We use the nebular regions defined in \citet{groves2023} as a mask and use this to identify each pixel as containing mostly emission from diffuse or nebular regions. We note that the physical resolution is $\approx 50$ times more coarse than our JWST observations of NGC~6822 and WLM, and so the delineation between ``inside" and ``outside" \hplus\ regions (\S\ref{subsec:7_21}) is not as clean for PHANGS \citep{egorov2023, sutter2024}. We sample the 21-cm intensities and corresponding \ion{H}{1} column densities at their native resolution, which is much coarser than the other data (typically $\gtrsim 10\arcsec \sim 1$~kpc).

The properties of these 19 galaxies are reported in \citet{leroy2021,emsellem2022,lee2023}. In our comparisons, we highlight the two lowest mass targets within these $19$ galaxies, which offer an intermediate case between the more massive typical PHANGS targets and our dwarf galaxy targets. IC~5332 is a low mass spiral with $\log_{10}~M_\star/M_\odot=9.67$, $\log_{10}~\mathrm{SFR}/(M_\odot~\mathrm{yr}^{-1})=-0.39$, and $12+\log_{10}\mathrm{O/H}=8.30$ dex at $r_\mathrm{eff}$ while NGC~5068 is a barred low mass spiral with $\log_{10}~M_\star/M_\odot=9.40$, $\log_{10}~\mathrm{SFR}/(M_\odot~\mathrm{yr}^{-1})=-0.56$, and $12+\log_{10}\mathrm{O/H}=8.32$ dex at $r_\mathrm{eff}$ \citep[see][]{leroy2021,groves2023}.

\subsubsection{The Magellanic Clouds}\label{subsec:mc}

At 49.6~kpc \citep{pietrzynski2019} and 62~kpc \citep{de-grijs2015, scowcroft2016}, the Large and Small Magellanic Clouds are by far the closest low mass star-forming galaxies. Their proximity means that previous-generation IR telescopes match the physical resolution achieved by JWST in our more distant Local Group targets. We compare our JWST F770W and F2100W measurements to \textit{Spitzer} 8~$\mu$m and 24~$\mu$m measurements of the LMC from SAGE \citep{meixner2006} and the SMC from SAGE-SMC \citep{gordon2011}. We use the publicly available maps whose processing is described in the survey papers, with additional processing as described in \citet{chastenet2019}. \citet{chastenet2019} conducted a careful background subtraction and beam matched all data at the $\approx 6\arcsec$ resolution of \textit{Spitzer}'s MIPS 24~\um\ filter ($\approx 1.4$~pc and $1.8$~pc at the distance to the LMC and SMC). To identify which regions of the galaxy correspond to star-forming \ion{H}{2} regions similar to our target fields, we use H$\alpha$ images of the LMC and SMC from the Southern H$\alpha$ Sky Survey Atlas \citep[SHASSA;][]{gaustad2001}, which has $48\arcsec$ ($\approx 12$~pc and $14$~pc) pixels.

\section{Results} \label{sec:results}

Figures~\ref{fig:rgb_6822_1} through \ref{fig:rgb_wlm} show our new JWST F770W and F2100W images. We detected extended emission in both filters in all four NGC~6822 fields. Field 3 shows a striking shell structure that resembles the infrared bubbles identified in the SMC \citep{ji2012} and Milky Way \citep{churchwell2004, churchwell2006, povich2007, watson2009, simpson2012, jayasinghe2019}. Meanwhile, Fields 2 and 4 both show compact, bright MIR emission that is coincident with the brightest CO emission in the galaxy, perhaps consistent with an early evolutionary state \citep[][]{schruba2017}.

The morphology of MIR emission in WLM (Fig.~\ref{fig:rgb_wlm}) appears fainter and much less extended than that in NGC~6822. This likely reflects WLM's lower metallicity, $Z \approx 0.13~Z_\odot$ compared to $Z \approx 0.2~Z_\odot$ in NGC~6822. The striking difference in MIR morphology may reflect that the PAH abundance and dust-to-gas ratio drop sharply over this metallicity range \citep[see reviews in][]{galliano2018,li2020,hunter2024}.

\begin{deluxetable*}{lccccc}
\tablecolumns{6}
 \tablecaption{Contributions of point sources to overall MIR emission}
    \tablehead{
         \colhead{Quantity} & 
         \colhead{NGC6822 \#1} &
         \colhead{NGC6822 \#2} &
         \colhead{NGC6822 \#3} &
         \colhead{NGC6822 \#4} &
         \colhead{WLM}
         }        
\startdata
\cutinhead{Mean intensity of hand masked bright point sources $\left< I_{\rm \nu}^{\rm ptsrc} \right>_{\rm reg}$ [$\times 10^{-3}$ MJy sr$^{-1}$]}
$\left< I_{\rm F770W}^{\rm ptsrc} \right>_{\rm reg}$ & 10 & 44 & 8 & 20 & 5\\ 
$\left< I_{\rm F2100W}^{\rm ptsrc} \right>_{\rm reg}$ & 5 & 22 & 10 & 5 & 5 \\ 
\cutinhead{Mean intensity of unmasked, fainter point sources $\left< I_{\rm \nu}^{\rm ptsrc} \right>_{\rm reg}$ [$\times 10^{-3}$ MJy sr$^{-1}$]} 
$\left< I_{\rm F770W}^{\rm ptsrc} \right>_{\rm reg}$  & 37 & 46 & 18 & 41 & 30 \\ 
$\left< I_{\rm F2100W}^{\rm ptsrc} \right>_{\rm reg}$  & 7 & 51 & 35 & 44 & 77 \\ 
\cutinhead{Fractional contribution of fainter point sources to total flux in field [\%]} 
$f_\mathrm{ptsrc}^\mathrm{F770W}$ & 19 & 11 & 14 & 14 & 41 \\ 
$f_\mathrm{ptsrc}^\mathrm{F2100W}$ & 1 & 2 & 6 & 3 & 24 \\ 
\cutinhead{Fraction of total MIR flux coming from CO-detected pixels  [\%]} 
$f^{\mathrm{F770W}}_{\mathrm{CO\ det.}}$ & 39.9 & 15.4 & 68.4 & 8.5 & 40.9 \\
$f^{\mathrm{F2100W}}_{\mathrm{CO\ det.}}$ & 56.4 & 19.2 & 85.5 & 7.4 & 72.2 \\
 \enddata
\tablecomments{
Fractional contributions of MIR point sources to the total emission in each JWST field $f_\mathrm{ptsrc}$ (Sec.~\ref{subsec:ptsrc}) at native resolution, and fraction of the total MIR flux coming from CO-detected pixels $f_\mathrm{CO~det.}$ at $2''$ resolution. $\left< I_{ \nu}^{\rm ptsrc} \right>_{\rm reg}$ and $f_\mathrm{ptsrc}$ are defined in Equations~\ref{eq:fptsrc2} and~\ref{eq:fptsrc}, respectively. Pixels with surface brightness $\leq 0$~MJy~sr$^{-1}$ were excluded from the calculations.}
\label{tab:ptsrc}
\end{deluxetable*}

In more detail, the images show: 
\begin{enumerate}
\item Though the F770W and F2100W images appear overall similar, the median \iratio\ color varies from field to field and appears lower in regions with the bright H$\alpha$ emission. We measure these variations and discuss their implications for PAH abundance variations in \S \ref{subsec:7_21}.

\item In NGC~6822, bright, compact dust emission appears associated with detected CO emission. We also see extended, lower intensity MIR dust emission surrounding these bright regions. This likely reflects some combination of CO-faint molecular gas, cold atomic gas, and the impact of dust heating by the nearby massive stars. We compare CO and MIR emission in \S \ref{sec:mir_co}.

\item Fainter diffuse MIR emission located far from the CO detections is visible in all four NGC~6822 fields. This may be dust associated with atomic gas clouds. Given the pervasive nature of the \ion{H}{1} it is just as remarkable that we do \textit{not} detect MIR emission along each line of sight, implying that much of the diffuse ISM may have low abundance of dust and PAHs. We compare MIR emission, \ion{H}{1}, and total gas column, $N({\rm H})$ in \S \ref{sec:mir_gas}.

\end{enumerate}

In addition to the dust, many point sources are visible in the images. We discuss their effect on our measurements in \S \ref{subsec:ptsrc}.

\subsection{Impact of MIR point sources} 
\label{subsec:ptsrc}

Our images show a large number of point sources. Many of these are MIR-bright or foreground stars, including young stellar objects \citep[YSOs; e.g.][]{jones2023, nally2024}. Some may be young clusters. There are also a handful of background galaxies. Our analysis focuses on dust emission, so that stars and background galaxies represent contaminants. To deal with these, we first mask bright sources that appear unrelated to our target galaxies, either because they are clearly background galaxies or appear to be foreground stars. We do this by hand.

After masking the bright sources, numerous fainter point sources remain. We attempted to identify and mask these based on F770W/F2100W colors, but this approach proved unreliable and we have deferred it to future work. We choose not to mask these fainter sources because distinguishing them from the extended, structured diffuse MIR emission that is the focus of our analysis presents significant challenges. Instead we use blank sky regions in each field to estimate the mean surface brightness due to faint point sources. We manually select regions in each field that show no diffuse MIR emission\footnote{We assume that the surface density of such sources is roughly uniform across each field. This should be justified for any population of older stars, as the fields are small compared to the full size of each target (e.g., Fig.~\ref{fig:overview}). However, these calculations will not reflect the contribution of YSOs or other sources selectively associated with the star-forming complexes themselves.}. Then we run \texttt{SExtractor} \citep{bertin1996} on each image. Within each control region we calculate the mean surface brightness of point sources over the whole region via:
\begin{equation}\label{eq:fptsrc2}
    \left< I_{\nu}^{\mathrm{ptsrc}}\right>_{\rm reg} \equiv \frac{1}{N_\mathrm{pix,reg}} \sum^{N_\mathrm{pix,reg}}I_\nu M_\mathrm{ptsrc},
\end{equation}
where $M_\mathrm{ptsrc}$ is a mask that is 1.0 for pixels in \texttt{SExtractor}-identified point sources and 0.0 otherwise. $N_{\rm pix,reg}$ refers to the total number of pixels in the region. Therefore, $\left< I_{\nu}^{\mathrm{ptsrc}}\right>_{\rm reg}$ is the surface brightness of emission associated with point sources averaged over the entire control region.

From $\left< I_{\nu}^{\mathrm{ptsrc}}\right>_{\rm reg}$, we estimate the fractional contribution of emission from point sources $f_\mathrm{ptsrc}$ to the overall emission in each field via
\begin{equation}\label{eq:fptsrc}
    f_\mathrm{ptsrc} = \frac{N_\mathrm{pix,map} \left< I_{\rm \nu}^{\rm ptsrc} \right>_{\rm reg} }{\sum^{N_\mathrm{pix,map}} I_\nu},
\end{equation}
where the sum now goes over each whole map, with $I_\nu$ the surface brightness of each pixel, and $N_\mathrm{pix,map}$ the number of pixels in the whole map. 

Both the mean intensities and the fractional contributions of point sources to the flux are reported in Table~\ref{tab:ptsrc}. Averaged over large areas, the mean intensities of point sources are comparable to or less than the noise in the images (Table~\ref{tab:gal}). Thus we expect them to have little impact on our analysis of the MIR color (\S \ref{subsec:7_21}) or comparison to CO (\S \ref{sec:mir_co}) where we focus on MIR intensities $\gtrsim 0.03$~MJy~sr$^{-1}$. They may affect our comparison to large-scale gas emission in \S \ref{sec:mir_gas}. 

The faint point sources contribute $1{-}19$\% of the total MIR flux in the NGC~6822 fields and $24{-}41\%$ in WLM. The higher values in WLM reflect that because of the faint, limited extended emission, point sources contribute fractionally more to WLM. The fact that $f_\mathrm{ptsrc}^\mathrm{F770W}$ is always larger than $f_\mathrm{ptsrc}^\mathrm{F2100W}$ reflects the higher sensitivity of F770W to emission from stars, which emit preferentially at shorter wavelengths.

\subsection{F770W vs. F2100W emission}\label{subsec:7_21}

\begin{figure*}
\begin{center}
\includegraphics[width=0.95\textwidth]{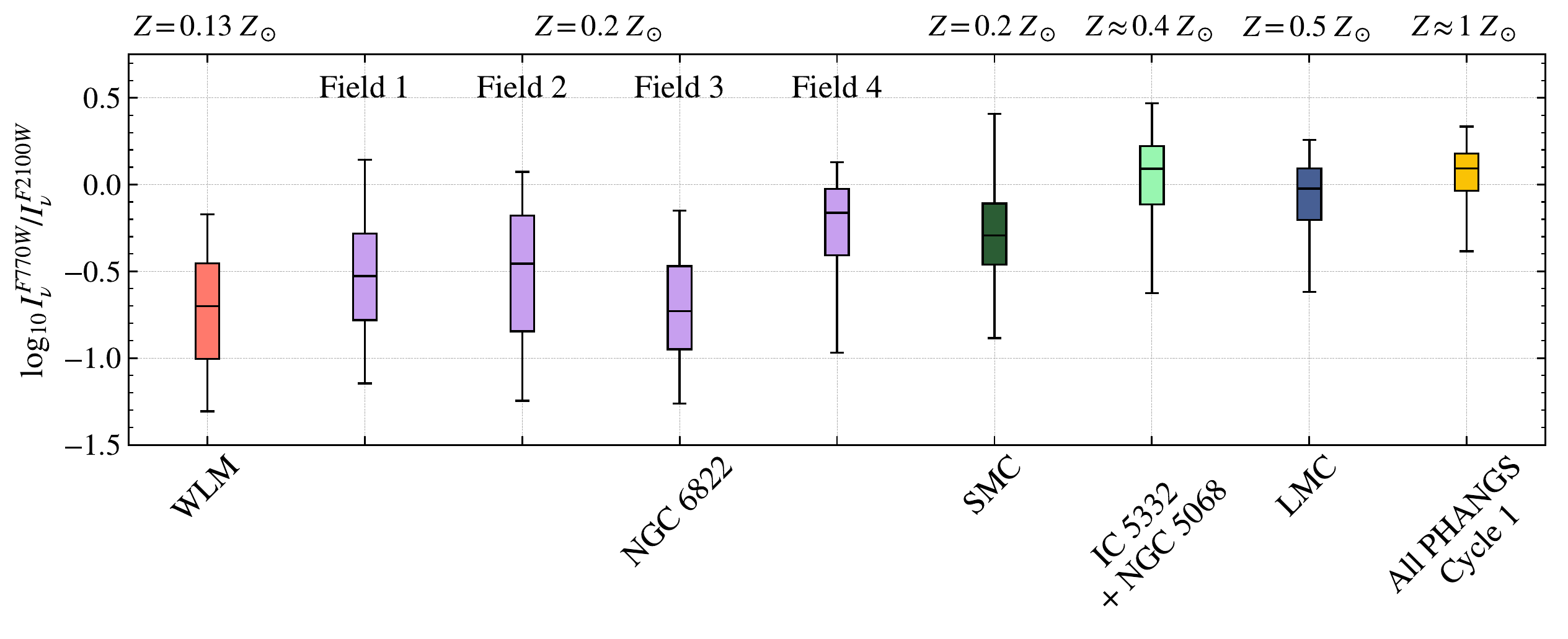}
\includegraphics[width=0.95\textwidth]{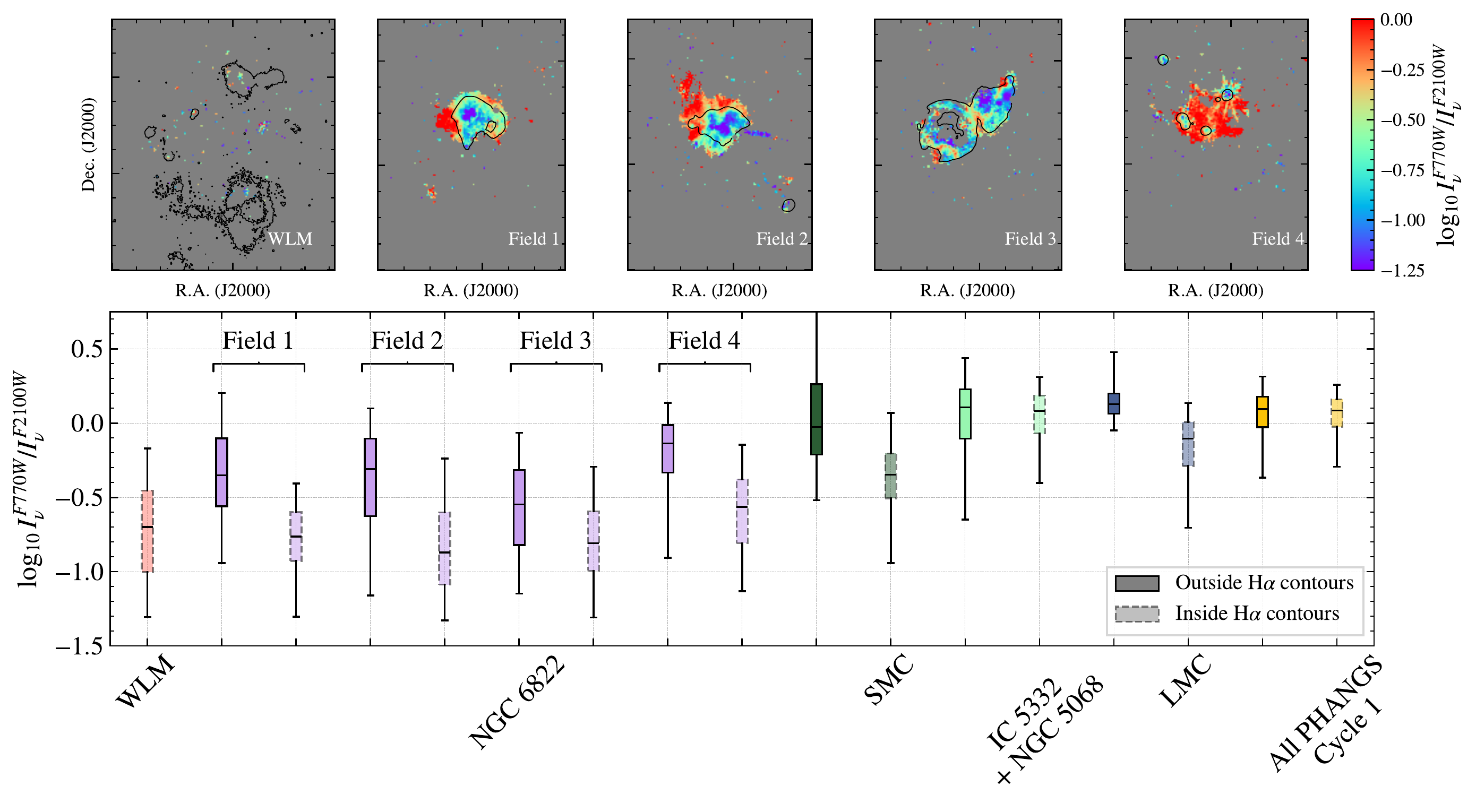}
\caption{\textit{Top:} Box and whisker plots of $\log_{10}$\iratio\ for pixels with S/N$>5$ in both filters for each target field and comparison data set. The boxes show the 16th, 50th, and 84th percentiles, while the whiskers show the 5th and 95th percentiles. The NGC~6822 and WLM distributions are from $0.9\arcsec\approx 2-4$~pc resolution maps. We compare with the Magellanic Clouds (at $6\arcsec\approx 1.5-2$~pc resolution) and PHANGS Cycle 1 galaxies at $\approx 100$~pc resolution, specifically highlighting the two dwarf galaxies in that sample, IC~5332 and NGC~5068. Metallicity increases from left to right (labeled at the very top of the figure). 
The \iratio\ drops below $Z\approx 0.2~Z_\odot$ and NGC~6822 shows significant variations between fields.
\textit{Middle:} \iratio\ ratio maps for WLM and NGC~6822. The black contour shows an H$\alpha$ intensity of $1\times 10^{-14}$~erg~s$^{-1}$~cm$^{-2}$~arcsec$^{-2}$, which by eye corresponds to boundaries of bright, compact emission, and a reasonable proxy for the extent of the \hplus\ regions. These H$\alpha$-bright regions also correspond to depressions in \iratio\ within the region. \textit{Bottom:} As the top row, but now separating data sets into H$\alpha$-bright regions and other emission (WLM lacks enough faint emission to separate). The H$\alpha$ bright regions show lower \iratio\ at all metallicities. Note that the inside/outside \hplus\ region distinction is coarse in PHANGS as the regions are not well-resolved (\S\ref{subsec:phangs} and \ref{subsec:7_21}). The values for the top and bottom row are provided in Table~\ref{tab:stats}.
\label{fig:violins}}
\end{center}
\end{figure*}

In the neutral ISM of solar-metallicity galaxies, the F770W filter captures mainly emission associated with the 7.7~\um\ PAH feature \citep[e.g.,][]{smith2007,whitcomb2023a,chown2025}. Meanwhile, continuum emission dominates the F2100W filter. At moderate radiation fields, both bands capture stochastically heated emission \citep[][]{draine2007a}. To first order, \iratio\ and the similar ratio $I_\nu^\mathrm{IRAC~8~\mu m}/I_\nu^\mathrm{MIPS~24~\mu m}$ measured by \textit{Spitzer} has often been taken to track the abundance of PAHs relative to larger grains \citep[e.g.,][]{engelbracht2005,gordon2008,li2020,chastenet2023,sutter2024}. We calculate \iratio\ for all pixels detected at SNR~$>5.0$ in both filters at $0.9''$ resolution and plot the distribution of this ratio in Fig.~\ref{fig:violins}. This figure also shows results for the LMC, SMC, and PHANGS targets.
Table~\ref{tab:stats} reports the median and scatter for each target. 

Fig.~\ref{fig:violins} shows a drop in the median \iratio\ for metallicities below $Z=0.4~Z_\odot$; i.e., NGC~6822, WLM, and the SMC all show lower \iratio\ or 8/24~$\mu$m compared to the LMC and the more massive PHANGS targets. 
Compared to the median PHANGS ratio $\log_{10} \iratio \approx 0.09$ or $\iratio \approx 1.23$, the NGC~6822 fields have on average a $0.58$~dex lower ratio, with the most extreme fields, NGC~6822 Field 3 and WLM, reaching $> 0.8$~dex, or $>6\times$, lower.

This correlation between \iratio\ and metallicity agrees with extensive literature work \citep[e.g.,][]{remy-ruyer2015,galliano2018,li2020}. Many \textit{Spitzer} observations indicated low PAH abundances below metallicity of $12 + \log_{10} {\rm O/H} \approx 8.2$, or $Z \approx 0.3 Z_\odot$ \citep[e.g.,][]{ engelbracht2005,cannon2006,jackson2006,draine2007,gordon2008}, and the SMC ($0.2~Z_\odot$) has lower PAH abundance than the LMC ($0.5~Z_\odot$) \citep{russell1992, chastenet2019}. It remains an open question whether the PAH abundance steadily declines as a function of metallicity \citep[e.g.,][]{galliano2018}, drops sharply at a specific $Z$ \citep[e.g.,][]{li2020,whitcomb2024}, or depends on a third parameter such as specific star formation rate \citep[][]{remy-ruyer2015}. 

Our observations demonstrate that the individual star-forming complexes at $Z=0.2~Z_\odot$ and $Z=0.13~Z_\odot$ show low but varied \iratio. We note that on their own these observations do not capture the transition from high to low metallicity, which would require a sample that better represents the galaxy metallicity distribution. We find significant scatter in the \iratio\ ratios between fields, with $0.56$~dex, or a factor of $3.6$, range among the median \iratio\ for different regions in NGC6822. At lower resolution, e.g., the 50{-}100~pc typical of PHANGS observations, each region will be a single resolution element and this would manifest as point-to-point scatter.

We also see scatter in \iratio\ within each region, and some of this appears to be physical in nature. The middle row in Figure \ref{fig:violins} shows the \iratio\ ratio map for each field. These show regular patterns, including regions of depressed \iratio\ (i.e., blue regions) within each field. These regions with low \iratio\ correspond to the brightest H$\alpha$-emitting regions in our targets. To show this, we plot contours of H$\alpha$ intensity in each field, with the level chosen by eye to correspond to the bright, compact H$\alpha$ emission.\footnote{A contour at $1\times10^{-14}$~erg~s$^{-1}$~cm$^{-2}$~arcsec$^{-2}$ captures the bright, compact H$\alpha$ emission in NGC~6822. Note that this contour is higher than the one used to separate the Magellanic Clouds into \ion{H}{2} regions and neutral gas by \citet{chastenet2019}.} In all four NGC~6822 fields, the H$\alpha$-bright regions correspond to the regions in the map with low \iratio.

The bottom row of Fig.~\ref{fig:violins} and Table~\ref{tab:stats} show \iratio\ split into H$\alpha$-bright and H$\alpha$-faint regions. Across all metallicities, \iratio\ drops inside \hplus\ regions. The drop from outside to inside \hplus\ regions is $\approx 0.50$~dex for NGC~6822 as a whole, $\approx 0.32$~dex for the SMC, and $\approx 0.23$~dex for the LMC. A drop of $\sim 0.2$~dex, similar in magnitude to our result, has been shown in the lower resolution PHANGS data by \citet[][and \citealt{chastenet2023a} and \citealt{egorov2023}]{sutter2024} where each resolution element correspond to a whole star-forming region (as noted in \S\ref{subsec:phangs}).

This result also agrees well with the literature. Reduced PAH to small dust grain intensity ratios have also been found in infrared bubbles in the Milky Way \citep[e.g.][]{churchwell2006, povich2007} and the SMC \citep[e.g.][]{sandstrom2010, cui2024}. Direct evidence for the suppression of PAH abundance within \hplus\ regions (compared to nearby well-shielded molecular regions) has been shown with JWST mid-IR spectroscopy of the Orion Bar PDR \citep{chown2024}. Our measurements show that this suppression of \iratio\ within the ionized gas of \ion{H}{2} regions also occurs within each individual star forming complex in NGC~6822. This lowers the ratio, which is already suppressed due to the low metallicity of the target, even further within \ion{H}{2} regions.

In Appendix~\ref{sec:appendix_qpah} we connect our \iratio\ measurements to radiation field intensity, PAH abundance, and dust grain size distribution based on dust models. The overall reduced \iratio\ seen in the dwarfs compared to PHANGS is qualitatively consistent with reduced \qpah\ that is expected at low $Z$. \iratio\ depends nonlinearly on \qpah, $U$, $\gamma$\footnote{$\gamma$ is the fraction of dust mass exposed to radiation fields stronger than $U_\mathrm{min}$. It represents the fraction of dust illuminated by a power-law distribution of radiation intensities between $U_\mathrm{min}$ and $U_\mathrm{max}$.} and dust size distribution, which we do not have constraints on, and so quantitatively ascribing the measured drop in \iratio\ compared to PHANGS requires assumptions about these secondary parameters. We find that the values of \iratio\ seen in the dwarfs imply reasonable values of \qpah\ for a range of possible radiation field intensity and $\gamma$. 

\subsection{A sublinear power law scaling between 7.7~\um\ and 21~\um\ at high intensity}
\label{sec:sublin}

\begin{figure*}
\begin{center}
\includegraphics[width=\textwidth]{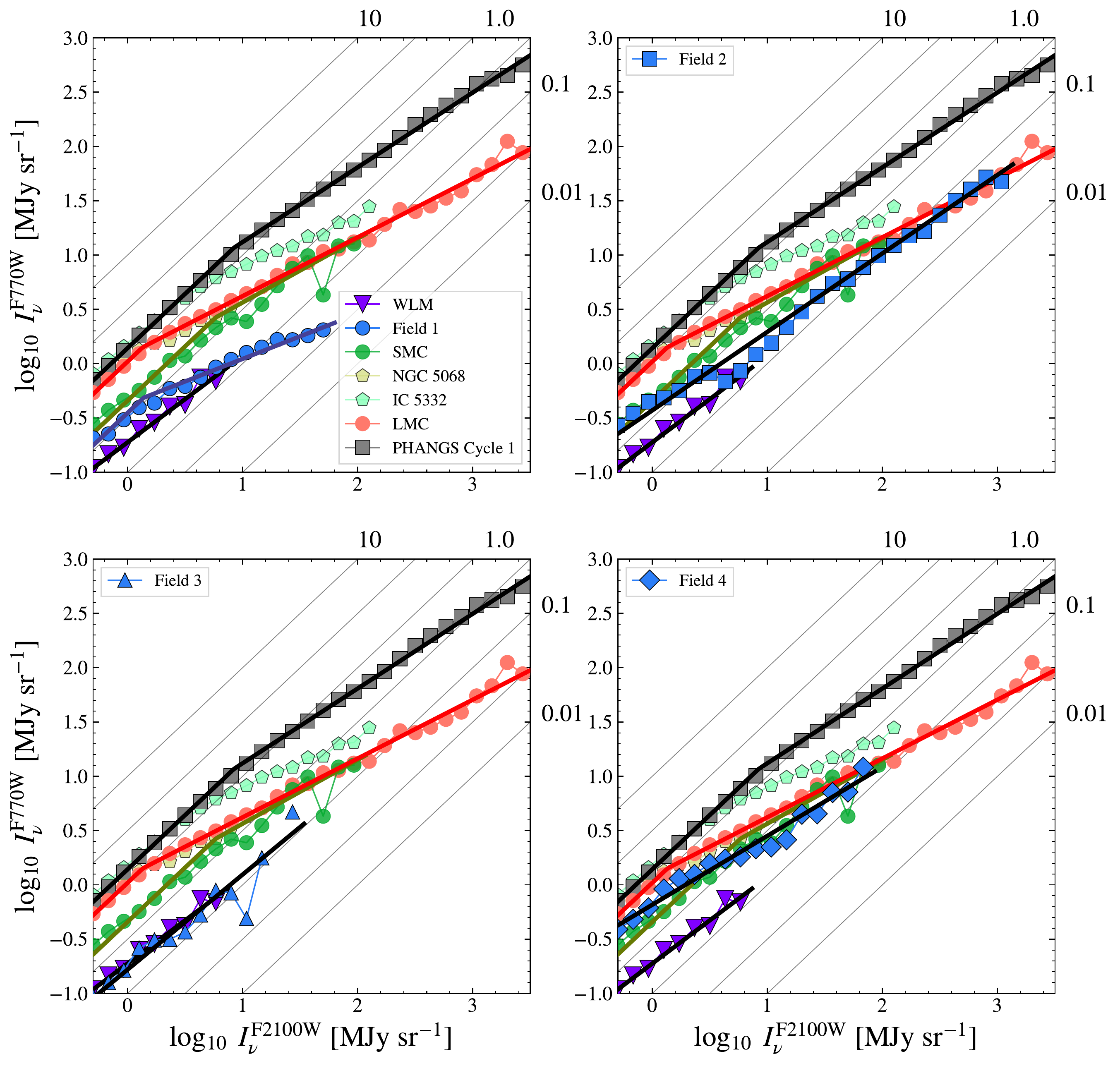}
\caption{Binned measurements of \ipah\ vs \idust\ at $0.9\arcsec$ ($\approx 2$--4~pc) resolution in NGC~6822 and WLM compared to galaxies with a range of metallicities. Each of the four panels is the same except for the NGC~6822 field shown.
For comparison we plot similar measurements for the LMC and SMC using \textit{Spitzer} IRAC 8~\um\ and MIPS 24~\um\ on the x-axis at the MIPS 24~\um\ resolution ($\approx 2$~pc); measurements from the full PHANGS Cycle 1 sample at coarser $\approx 100$~pc resolution; and the specific PHANGS measurements for the dwarf galaxies NGC~5068 and IC~5332. Diagonal gray lines show constant \iratio . Fits, either single or broken powerlaw fits appear as solid lines (\S~\ref{sec:sublin} and Table~\ref{tab:fits}). The individual data sets show vertical offsets that can be attributed to metallicity (\S \ref{subsec:7_21}, Fig. \ref{fig:7_vs_21_ratio}) but also a common shape, such that above $\idust \approx 1$~MJy~sr$^{-1}$ (i.e., $x=0$) all data sets show a sublinear power law, in which \ipah / \idust\ declines with increasing \idust . This reflects the suppression of \ipah\ relative to \idust\ within \ion{H}{2} regions due to PAH destruction in ionized gas.
\label{fig:7_vs_21}}
\end{center}
\end{figure*}

\begin{figure*}
\begin{center}
\includegraphics[width=\textwidth]{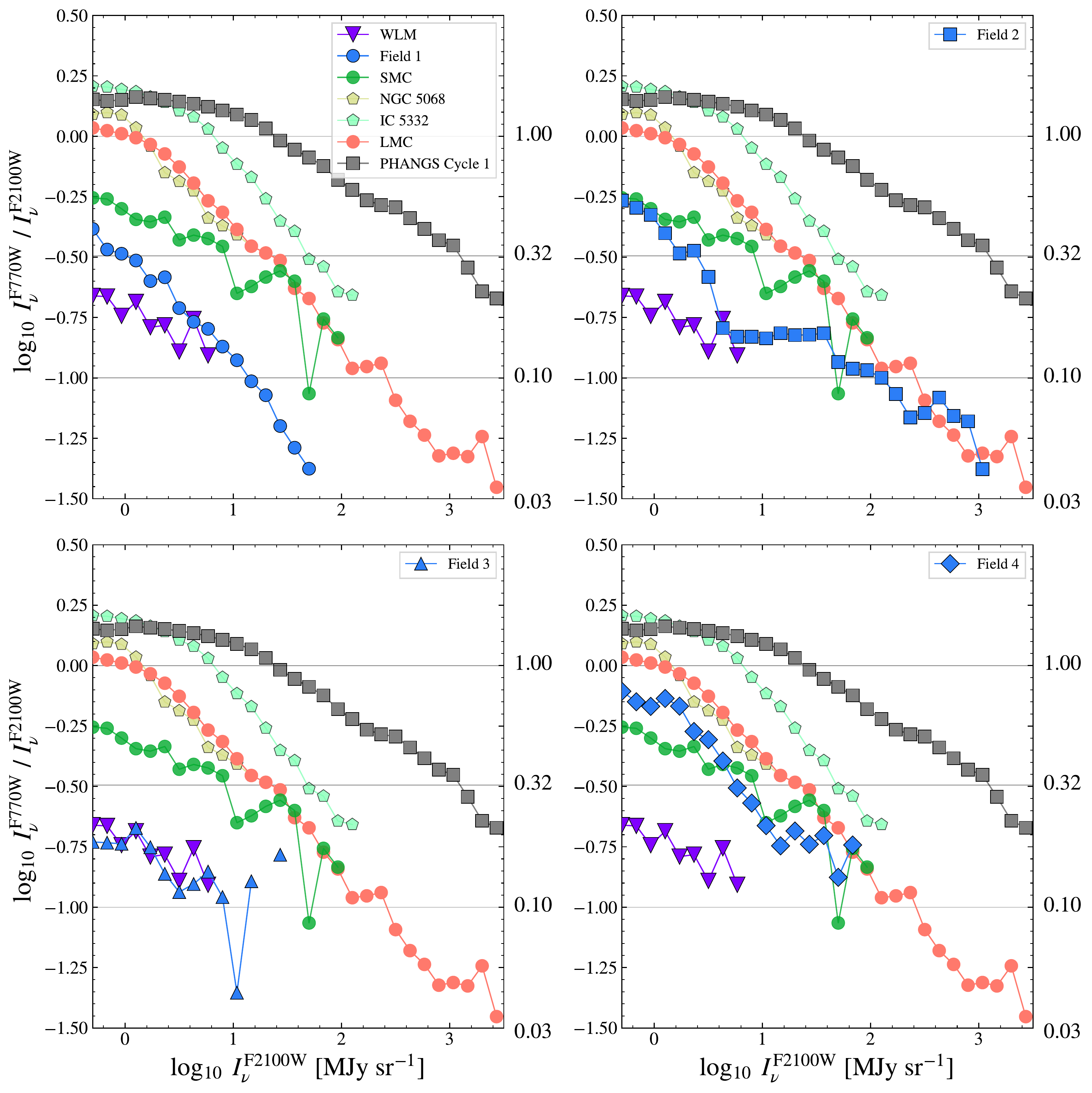}
\caption{The same measurements from Fig.~\ref{fig:7_vs_21} but with the y-axis expressed as a ratio. Horizontal lines indicate the constant ratios shown as diagonal lines in Fig.~\ref{fig:7_vs_21}. As in Fig.~\ref{fig:7_vs_21} the data show vertical offsets that can be attributed to metallicity and a steady decline in \ipah / \idust\  with increasing \idust\ at high \idust . This reflects the suppression of \ipah\ relative to \idust\ within \ion{H}{2} regions due to PAH destruction in ionized gas. \label{fig:7_vs_21_ratio}
}
\end{center}
\end{figure*}

\begin{deluxetable*}{lrrrrr}
 \tablecaption{Median and scatter of $\log_{10}$F770W/F2100W}
    \tablehead{
         \colhead{} & \colhead{} & \multicolumn{4}{c}{ $\log_{10}$(median F770W/F2100W) [dex]} \\
         \colhead{Name} & \colhead{$Z/Z_\odot$} & \colhead{Full area} & \colhead{Outside H$\alpha$ cont.} & \colhead{Inside H$\alpha$ cont.} & \colhead{Inside/outside} \\
         \colhead{(1)} & \colhead{(2)} & \colhead{(3)} & \colhead{(4)} & \colhead{(5)} & \colhead{(6)}}
\startdata
WLM & 0.13 & $-0.70 \pm 0.40$ & -- & -- & -- \\
NGC~6822 & 0.20 & $-0.49 \pm 0.47$ & $-0.30 \pm 0.35$ & $-0.80 \pm 0.30$ & $-0.50 \pm 0.47$ \\
... Field 1 & -- & $-0.53 \pm 0.37$ & $-0.35 \pm 0.34$ & $-0.76 \pm 0.24$ & $-0.41 \pm 0.42$ \\
... Field 2 & -- & $-0.47 \pm 0.48$ & $-0.32 \pm 0.36$ & $-0.87 \pm 0.35$ & $-0.55 \pm 0.51$ \\
... Field 3 & -- & $-0.73 \pm 0.35$ & $-0.55 \pm 0.37$ & $-0.81 \pm 0.29$ & $-0.25 \pm 0.47$ \\
... Field 4 & -- & $-0.17 \pm 0.25$ & $-0.14 \pm 0.22$ & $-0.56 \pm 0.30$ & $-0.42 \pm 0.37$ \\
SMC & 0.20 & $-0.29 \pm 0.26$ & $-0.03 \pm 0.32$ & $-0.35 \pm 0.22$ & $-0.32 \pm 0.39$ \\
PHANGS Dwarfs & 0.40 & $0.06 \pm 0.23$ & $0.08 \pm 0.22$ & $0.08 \pm 0.18$ & $-0.00 \pm 0.28$ \\
LMC & 0.50 & $-0.02 \pm 0.21$ & $0.13 \pm 0.10$ & $-0.10 \pm 0.20$ & $-0.23 \pm 0.22$ \\
PHANGS & 1 & $0.09 \pm 0.15$ & $0.09 \pm 0.14$ & $0.08 \pm 0.13$ & $-0.01 \pm 0.19$ \\
 \enddata
\tablecomments{(3) All pixels included. \\
(4) Including pixels outside the H$\alpha$ contour $I(\mathrm{H\alpha})=1\times 10^{-14}$~erg~s$^{-1}$~cm$^{-2}$~arcsec$^{-2}$ (to capture most of the H$\alpha$ emission) and $I(\mathrm{H\alpha})=1.5\times 10^{-15}$~erg~s$^{-1}$~cm$^{-2}$~arcsec$^{-2}$ for the Magellanic Clouds \citep[from][]{chastenet2019}. We split the PHANGS sample in a similar way using the catalog of \hplus\ regions (\S~\ref{subsec:phangs}).  
\\
(5) Including pixels within the H$\alpha$ contour. \\
(6) Ratio of column 5 to 4 (in $\log_{10}$ units).\\ 
}
\label{tab:stats}
\end{deluxetable*}

Regions with suppressed \iratio\ and bright H$\alpha$ coincide with the highest MIR intensities observed in our fields. While \iratio\ is suppressed in H$\alpha$-bright regions, these regions exhibit the highest absolute intensity values in our fields. Figure~\ref{fig:7_vs_21} illustrates this relationship through the median \ipah\ as a function of \idust, and Figure~\ref{fig:7_vs_21_ratio} demonstrates \iratio\ as a function of \idust.
One can see from these figures that at high $\idust$ (or high $\ipah$, not shown) the ratio \iratio\ drops as a function of increasing MIR brightness, manifesting as a sublinear power law relationship between $\ipah$ and $\idust$ at high intensities. 

The SMC, LMC, PHANGS dwarfs, and NGC~6822 all show a common behavior in Fig.~\ref{fig:7_vs_21}. At low \idust, the targets show approximately fixed \iratio\ leading to an approximately linear $\ipah{-}\idust$ relationship. Then above $\idust \approx 1{-}5$~MJy~sr$^{-1}$ they show declining \iratio\ with increasing \idust. The entire PHANGS data set shows similar behavior, but with the turnover from a fixed to declining ratio (or linear to sub-linear power law) at higher \idust\ (i.e., $x_b$ in Eq.~\ref{eq:broken}). We note that WLM does not show the same two-component behavior in our data, likely because we do not detect an extensive diffuse, fixed \iratio\ component.

This behavior can be expressed as a two-part power law,
\begin{equation}\label{eq:broken}
y=
\begin{cases}
    a + b_1(x-x_b)& \text{for } x \leq x_b \\
    a + b_2(x-x_b) & \text{for } x > x_b,
\end{cases}
\end{equation}
relating F770W intensity $y\equiv\log_{10} I_{\nu}^{\rm F770W}$ to $x\equiv\log_{10} I_{\nu}^{\rm F2100W}$. Because each of the binned relations in Fig.~\ref{fig:7_vs_21} has a component at low intensity (around 1~MJy~sr$^{-1}$) that is very close to a constant $\ipah/\idust$ ratio (i.e., a power law slope of 1.0), we fix $b_1=1.0$; this also makes it easier to compare with Figure~\ref{fig:violins} and Table~\ref{tab:stats}. Table~\ref{tab:fits} reports broken power law fits of this form for each target. Some regions are not well fit by a broken powerlaw (with $\leq 4$ bins below $x_b$, making that part of the fit unreliable), and so for these regions we fit to a powerlaw,
\begin{equation}\label{eq:line}
y= p + q(x-x_0),
\end{equation}
where $x_0\equiv \mathrm{median}(x)$. 

Fig.~\ref{fig:7_vs_21_ratio} resembles the relationship between $R_{\rm PAH}$ as a function of H$\alpha$ intensity \citep[][]{chastenet2019,sutter2024}. This makes sense given the well-established correlation between $\idust$ and H$\alpha$ intensity \citep[e.g.,][]{belfiore2023}. This, in turn, reflects the strong dust heating by UV radiation in and around \hplus\ regions and perhaps also the concentration of dust near \hplus\ regions. Then the break in the relationship and onset of declining \iratio\ corresponds to where \hplus\ regions dominate the emission along the line of sight, while the nearly fixed \iratio\ at lower intensity corresponds to the neutral ISM. Then the vertical offsets among sources in Fig.~\ref{fig:7_vs_21} just reflect metallicity dependence of \iratio. The higher \idust\ turnover ($x_b$) where the \iratio-\idust\ anti-correlation begins in the main PHANGS sample likely reflects the higher dust column densities associated with the diffuse ISM in those more massive, dust-rich galaxies \citep[e.g.,][]{pathak2024}.

\begin{deluxetable*}{lrrrrr}
 \tablecaption{Best-fit parameters of the F770W vs F2100W relationship modeled as a broken powerlaw (Eq.~\ref{eq:broken}), or single powerlaw (Eq.~\ref{eq:line}). %
 }
    \tablehead{
         \colhead{Name} & \colhead{$Z/Z_\odot$} & \colhead{$x_0$ [$\log_{10}$MJy~sr$^{-1}$]} & \colhead{$x_b$ [$\log_{10}$MJy~sr$^{-1}$]} & \colhead{$b_2$} & \colhead{$a$ [$\log_{10}$MJy~sr$^{-1}$]}}

\startdata
WLM & 0.13 & $0.167$ & -- & $0.796 \pm 0.207$ & $-0.592 \pm 0.133$ \\
NGC6822 & 0.20 & $1.300$ & -- & $0.741 \pm 0.113$ & $0.464 \pm 0.129$ \\
... Field 1 & -- & -- & $0.139 \pm 0.550$ & $0.416 \pm 0.329$ & $-0.314 \pm 0.466$ \\
... Field 2 & -- & $1.300$ & -- & $0.723 \pm 0.115$ & $0.509 \pm 0.130$ \\
... Field 3 & -- & $0.433$ & -- & $0.876 \pm 0.249$ & $-0.401 \pm 0.190$ \\
... Field 4 & -- & $0.700$ & -- & $0.634 \pm 0.147$ & $0.261 \pm 0.125$ \\
SMC & 0.20 & -- & $0.767 \pm 0.457$ & $0.589 \pm 0.266$ & $0.426 \pm 0.431$ \\
IC 5332 & 0.37 & -- & $0.407 \pm 0.451$ & $0.487 \pm 0.296$ & $0.593 \pm 0.435$ \\
NGC 5068 & 0.38 & $0.367$ & -- & $0.606 \pm 0.168$ & $0.227 \pm 0.123$ \\
LMC & 0.50 & -- & $0.135 \pm 0.445$ & $0.542 \pm 0.178$ & $0.154 \pm 0.407$ \\
PHANGS & 1.00 & -- & $0.926 \pm 0.478$ & $0.688 \pm 0.209$ & $1.069 \pm 0.467$ \\
 \enddata
\tablecomments{For the LMC and SMC, we replace F770W with IRAC 8~\um\ and F2100W with MIPS 24~\um. Fits were performed on binned measurements, incorporating uncertainties in x and y. For binned measurements, x uncertainties were set to half of the bin width. $\log_{10}x_0$ is the median of the x-values that go into the fit in the event that a single powerlaw provides a better fit, and in these cases, the model is Equation~\ref{eq:line}.
}
\label{tab:fits}
\end{deluxetable*}

\subsection{MIR and CO(2-1) emission} \label{sec:mir_co}

\begin{figure*}
\begin{center}
\includegraphics[width=0.495\textwidth]{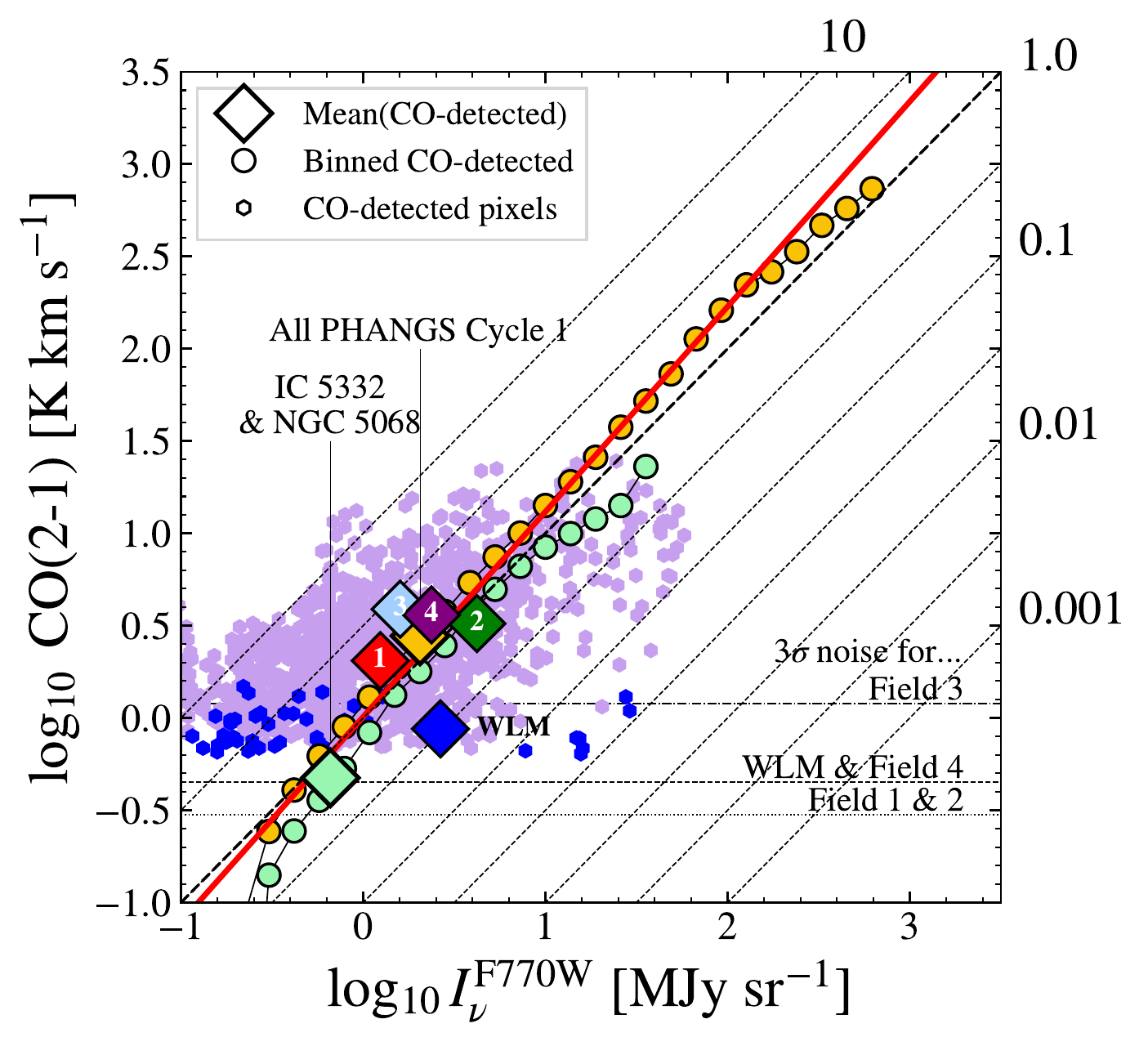}
\includegraphics[width=0.495\textwidth]{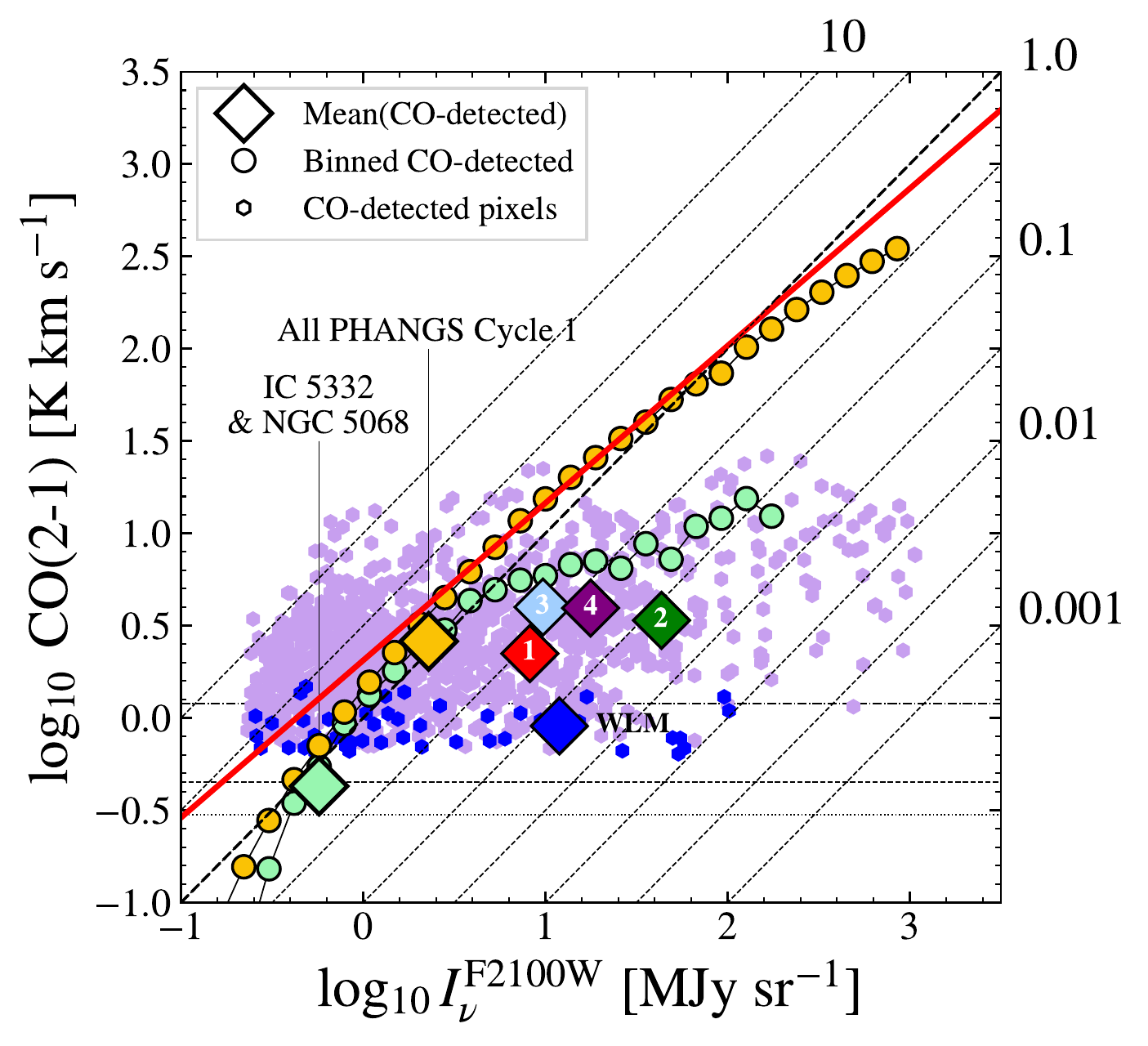}
\caption{CO(2-1) intensity as a function of \ipah\ (left) and \idust\ (right) in NGC~6822 (small, purple symbols) and WLM (small, light red symbols) for CO-detected pixels (at $\geq 3\sigma$) at $2\arcsec$ resolution. Diagonal lines indicate constant CO(2-1)/MIR ratios. Horizontal lines indicate the $3\sigma$ CO noise levels for our data (Section~\ref{subsec:data_alma}). 
Diamond symbols show the average over CO-detected pixels for each sample. Fields 1--4 and WLM are labeled on or next to their diamond symbols.  For comparison we show results for all PHANGS Cycle 1 galaxies (gold), and PHANGS dwarf galaxies (green). Solid lines show fits to these binned PHANGS data. While individual pixels and region-averages show some scatter, the CO to \ipah\ ratio for detected regions in NGC~6822 and WLM resembles that seen for more massive and more distant PHANGS galaxies. In contrast, the CO to \idust\ ratio appears lower in our targets compared to that observed in the more massive PHANGS targets, with NGC~6822 and WLM show $\approx 3-20\times$ fainter CO emission at fixed F2100W surface brightness (Fig. \ref{fig:co_mir_ratios}). CO/MIR ratios corresponding to the large diamonds are presented in Table~\ref{tab:gasdust1}.
\label{fig:co_vs_mir}}
\end{center}
\end{figure*}

\begin{figure*}
\begin{center}
\includegraphics[width=0.495\textwidth]{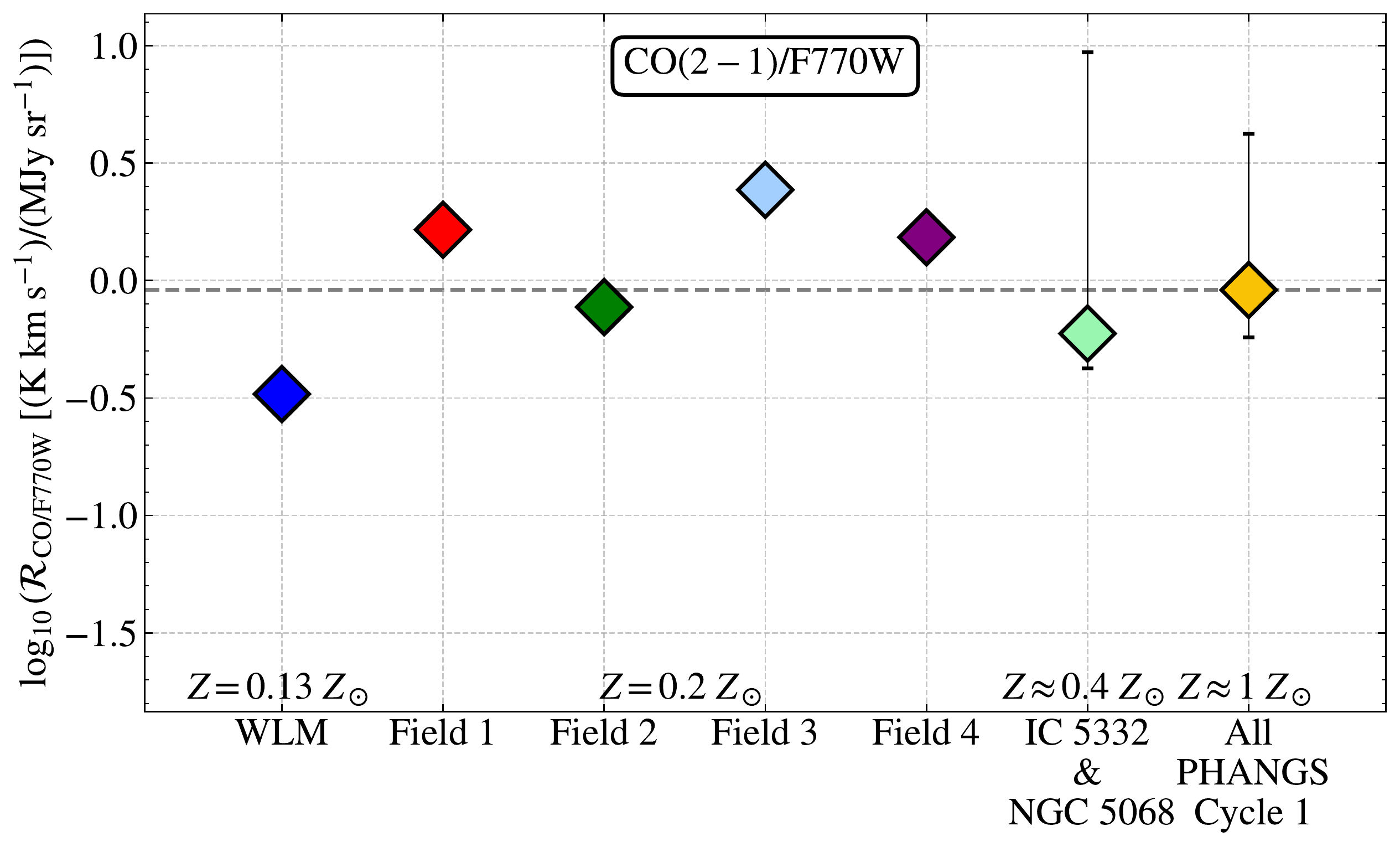}
\includegraphics[width=0.495\textwidth]{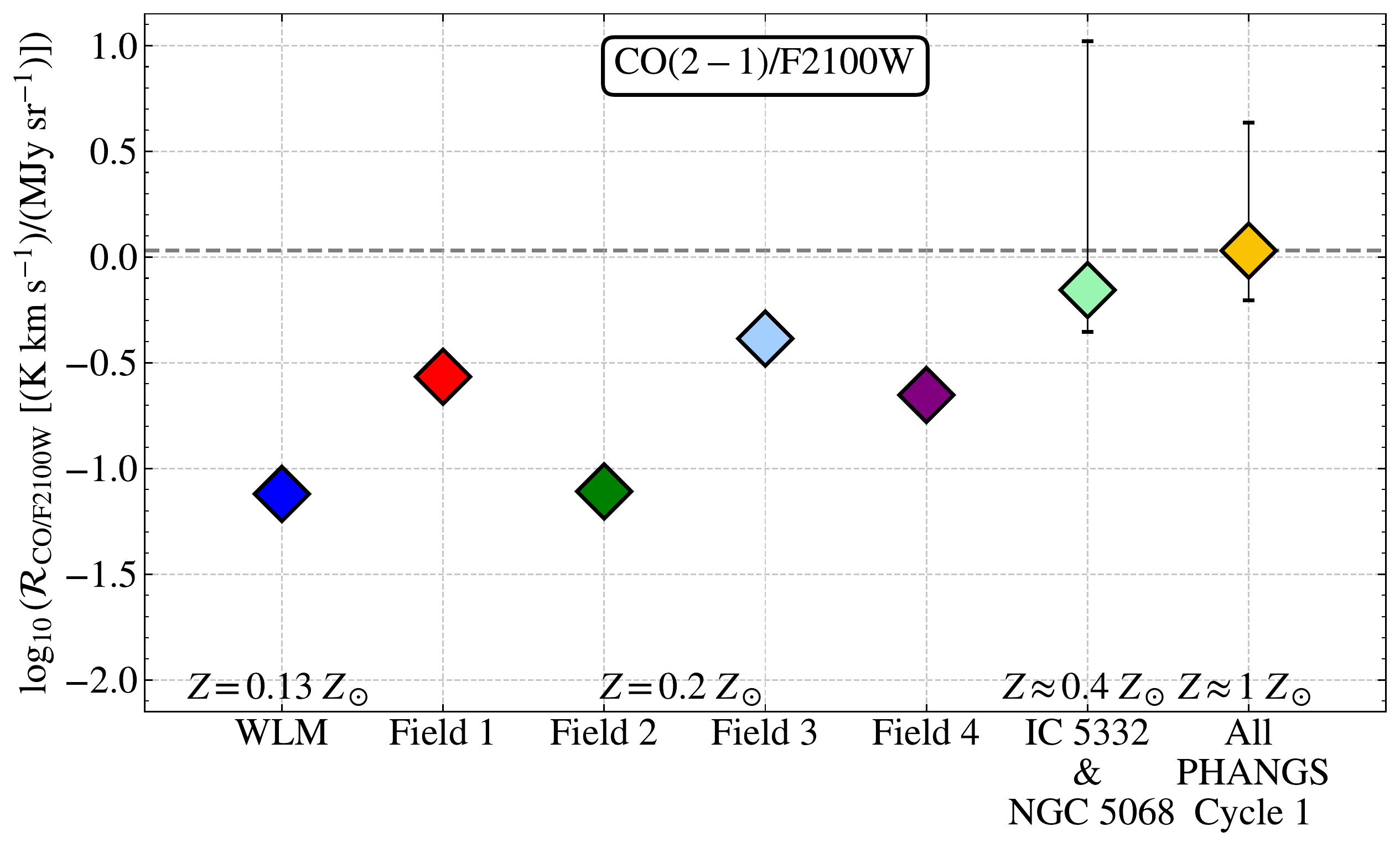}
\includegraphics[width=0.495\textwidth]{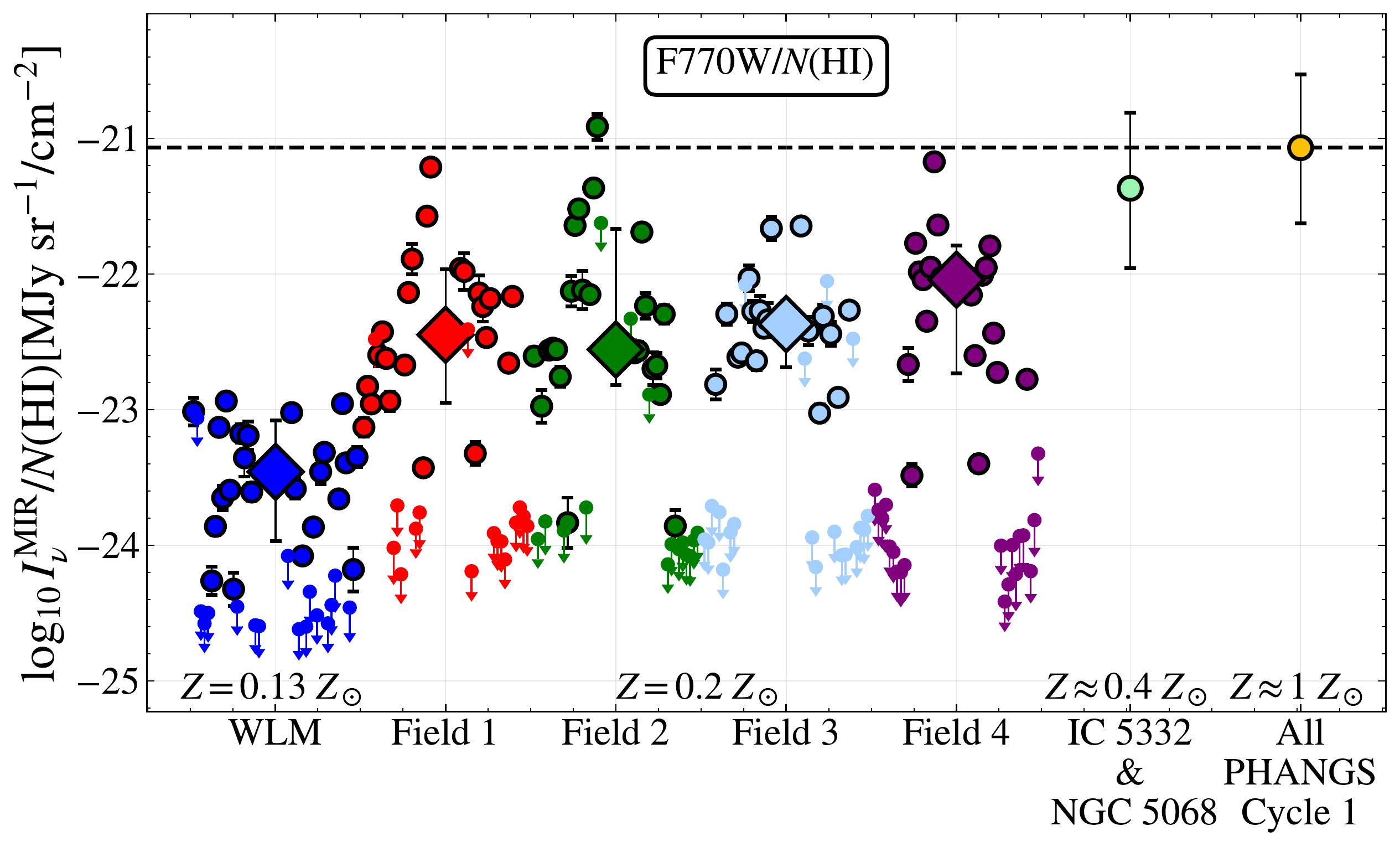}
\includegraphics[width=0.495\textwidth]{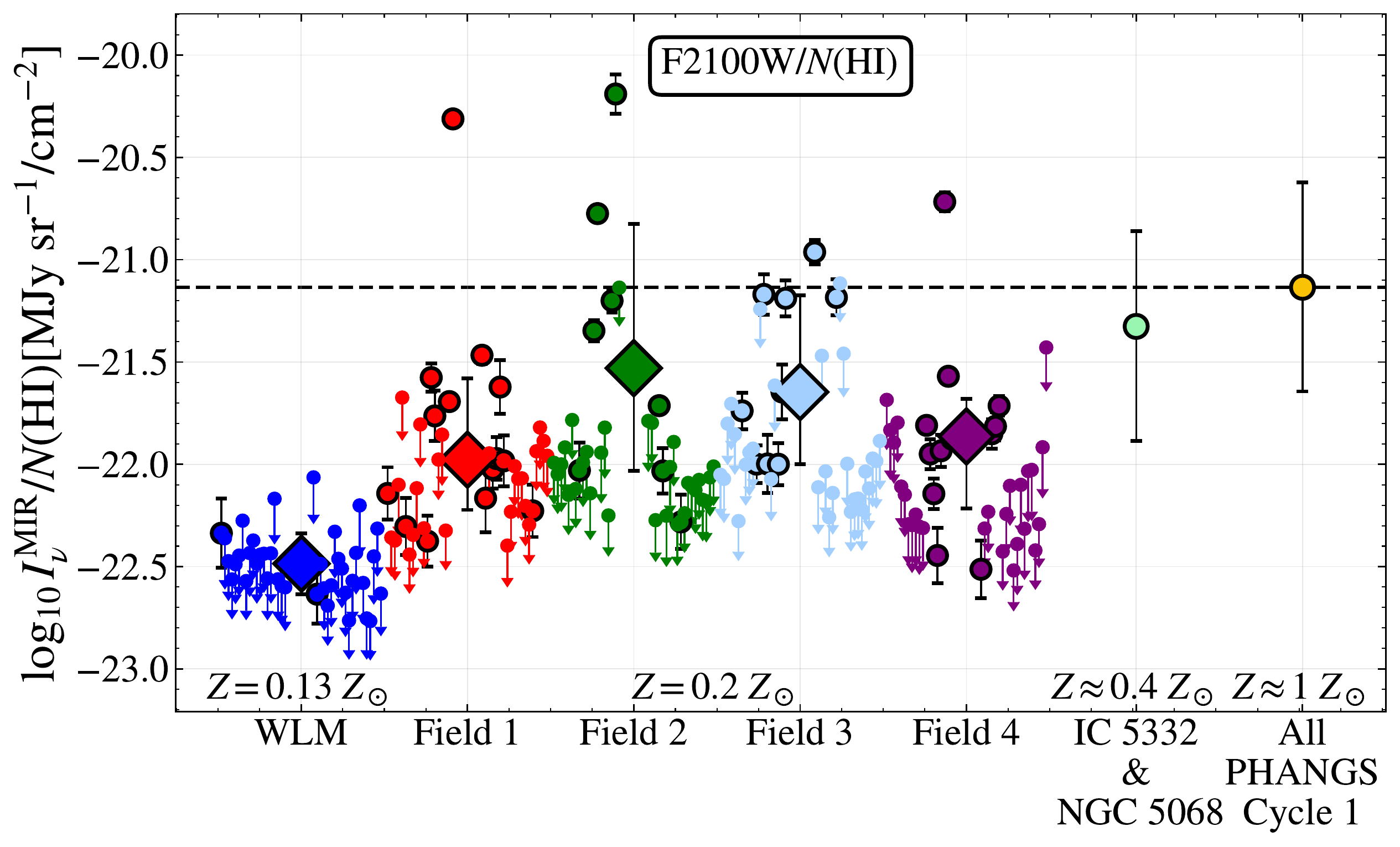}
\includegraphics[width=0.495\textwidth]{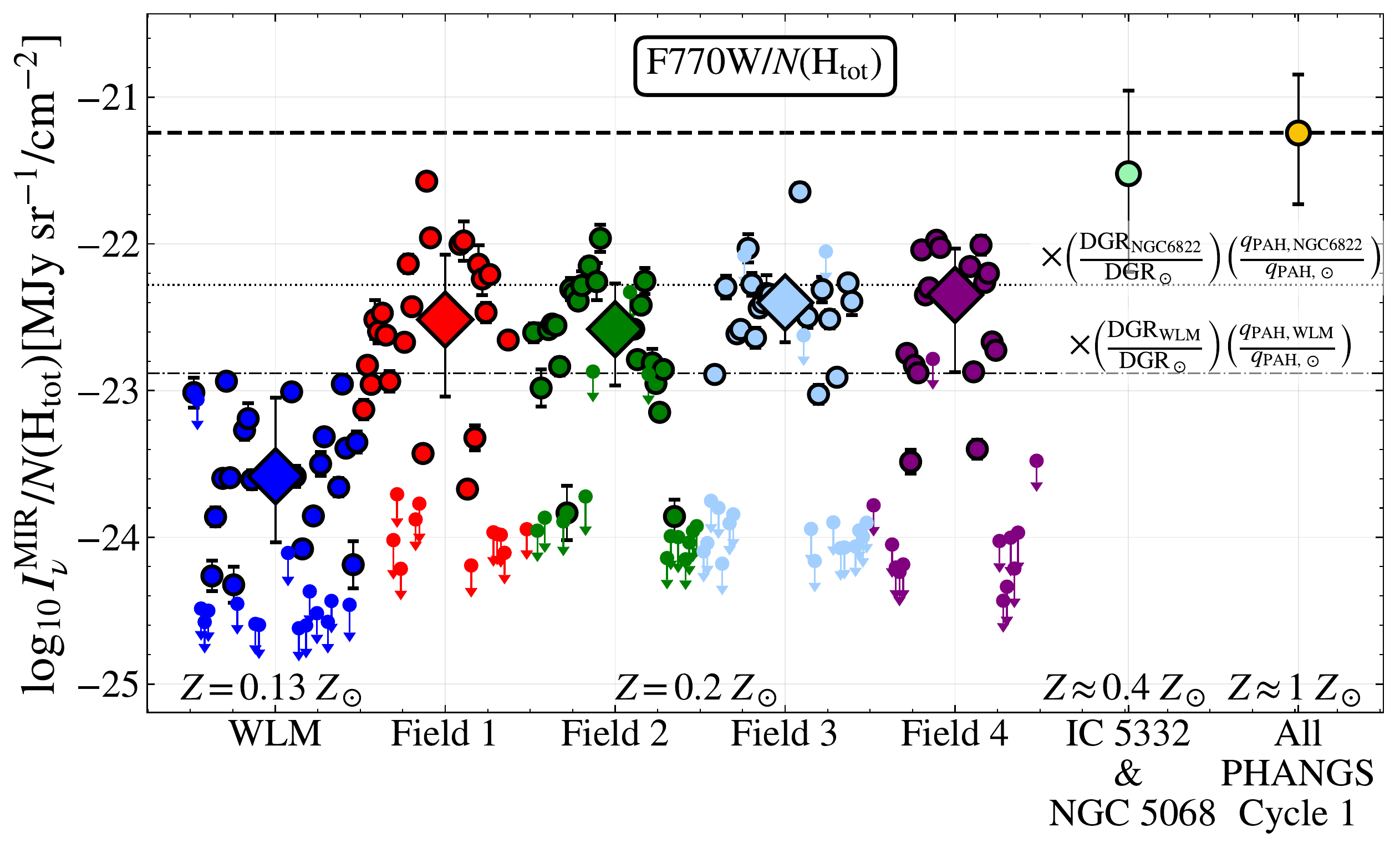}
\includegraphics[width=0.495\textwidth]{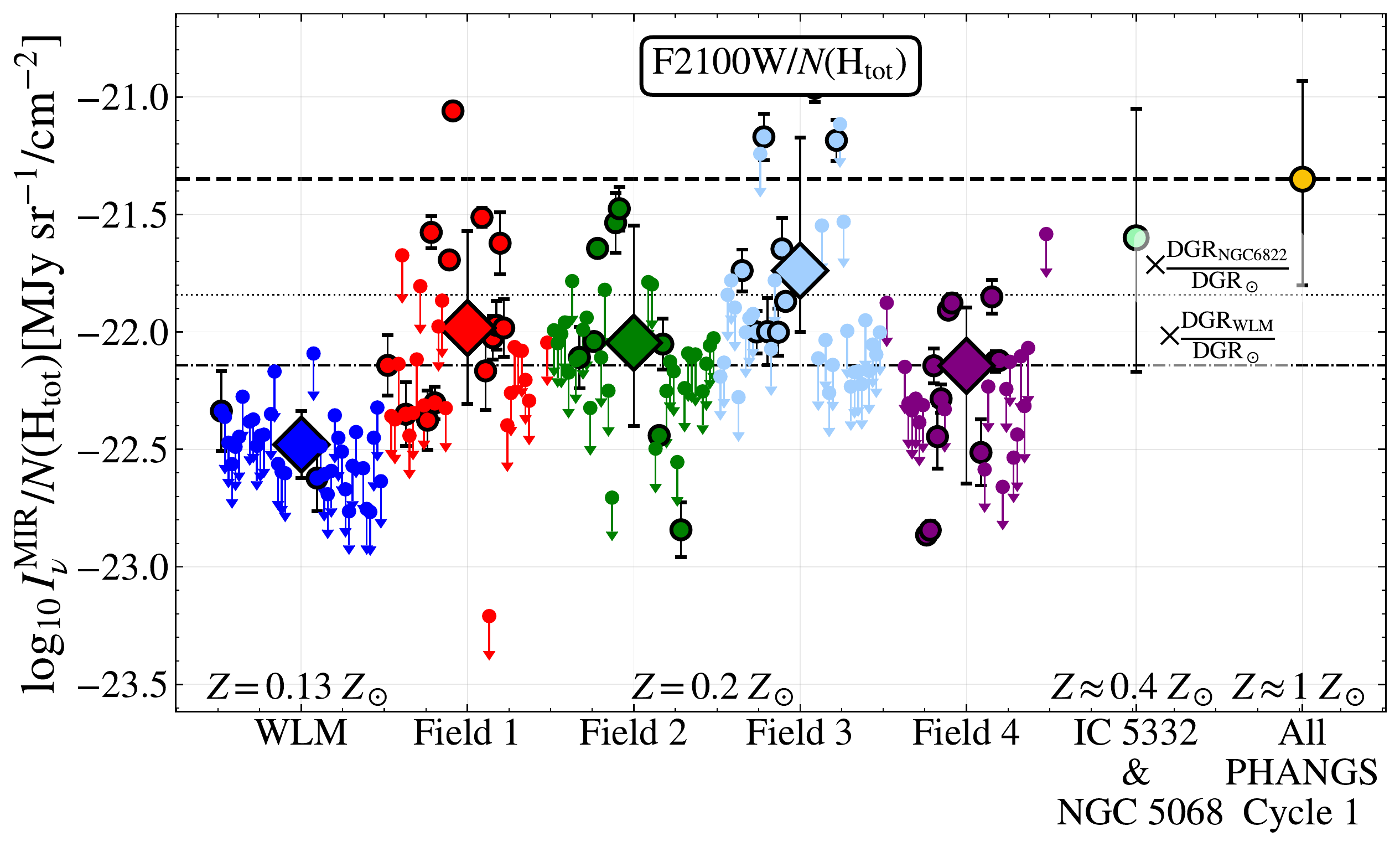}
\caption{\textbf{Ratios of MIR emission to gas tracers.} Top row: Average $I_\mathrm{CO(2-1)}$ divided by average \ipah\ (left) and \idust\ (right) (Equation~\ref{eq:r_comir}). The horizontal line indicates the median CO/MIR intensity ratio for PHANGS. The error bars show 16th-84th percentiles. The averages were calculated from CO and MIR maps at $2\arcsec$ resolution, and only including pixels where CO emission is detected and the MIR intensity has SNR~$>5$. Metallicity increases from left to right (but not to scale). NGC~6822 and WLM exhibit CO/\ipah\ ratios similar to those in PHANGS, while their CO/\idust\ ratios are lower by $0.5{-}1.5$~dex. The second and third rows show $I_\nu^\mathrm{MIR}/N(\mathrm{HI})$ and $I_\nu^\mathrm{MIR}/N(\mathrm{H_{tot}})$, calculated in $10''$ radius circular apertures that cover the fields (Fig.~\ref{fig:f770w_apertures}). The diamonds and error bars indicate the median and 16-84th percentile for apertures with detected MIR emission. WLM and NGC~6822 both show suppressed $I_\nu^\mathrm{MIR}/N(\mathrm{HI})$ and $I_\nu^\mathrm{MIR}/N(\mathrm{H_{tot}})$ compared to PHANGS. The dashed lines in the bottom panels show the expected ratios accounting for the lower dust to gas ratio and PAH abundance in our targets. These lower abundances appear to explain the typical $I_\nu^\mathrm{MIR}/N(\mathrm{H_{tot}})$ detected regions of NGC~6822, but upper limits and measurements for WLM suggest additional suppression of PAH and dust emission from the diffuse ISM. For the corresponding scatter plots see Fig.~\ref{fig:co_vs_mir} (top row) and Fig.~\ref{fig:mir_vs_nh_and_nhtot} (middle and bottom row). The per-field medians are shown in Table~\ref{tab:gasdust1}, and compared to PHANGS in Table~\ref{tab:gasdust2}.
\label{fig:co_mir_ratios}}
\end{center}
\end{figure*}

Figures~\ref{fig:rgb_6822_1} through \ref{fig:rgb_wlm} show bright MIR emission coincident with CO~(2-1) emission detected by ALMA, and a main goal of our project is to use the MIR to trace otherwise hard-to-observe gas (\S \ref{sec:intro}) in these complexes. To that end, Fig.~\ref{fig:co_vs_mir} shows the relationships between \ipah\ and CO~(2-1), and between \idust\ and CO~(2-1). The figure shows CO-detected pixels only, because unfortunately the lack of reliable short spacing data precludes a stacking analysis in the ALMA data for these galaxies \citep[e.g., see][]{neumann2023}. The square symbols show the mean CO-to-MIR ratio for each of our target fields. Then Fig.~\ref{fig:co_mir_ratios} shows these ratios as a function of metallicity for our target fields and comparison data. We quote the ratio of sums
\begin{equation}\label{eq:r_comir}
    \begin{split}
        \frac{\mathcal{R}_\mathrm{CO/X}}{\mathrm{(K~km~s^{-1})/(MJy~sr^{-1})}} \equiv & \left(\sum_{\rm CO-det.}\frac{I_\mathrm{CO(2-1)}}{\mathrm{K~km~s^{-1}}}\right) \\
        & \times \left(\sum_{\rm CO-det.}\frac{ I_\nu^X}{\mathrm{MJy~sr^{-1}}}\right)^{-1},
    \end{split}
\end{equation}
where $X$ is the MIR band (either F770W or F2100W). The sum runs over the pixels with detected CO~(2-1) emission, and we report the fraction of MIR emission associated with this selection in Table~\ref{tab:ptsrc}.

Fig.~\ref{fig:co_vs_mir} shows a large amount of scatter with many of the detected pixels from ALMA close to the noise limit in both galaxies. However, the mean CO-to-\ipah\ ratios for all four NGC~6822 fields and WLM resemble those found for the more massive galaxies by \citet{chown2025}. In fact, as Fig.~\ref{fig:co_mir_ratios} and Table~\ref{tab:gasdust1} show, we observe similar CO-to-\ipah\ ratios between the whole PHANGS sample, the PHANGS dwarf subset, or the fields in NGC~6822. WLM shows a modest $0.44$~dex depression with fainter CO relative to F770W ($\approx 3.4\times$ below PHANGS). On the other hand, the right panels in Figs.~\ref{fig:co_vs_mir} and \ref{fig:co_mir_ratios} show that the CO-to-\idust\ ratio does appear lower in our targets compared to more massive galaxies, with all four fields in NGC~6822 well below the CO-to-\idust\ ratio measured for PHANGS targets and WLM an order of magnitude lower (Tables~\ref{tab:gasdust1} and \ref{tab:gasdust2}). Overall, the CO/\idust\ ratio appears to show some metallicity dependence, with the lowest values for WLM, then NGC~6822, then the PHANGS dwarfs, then the full PHANGS sample.

Our results show that as metallicity decreases, CO and PAH emission decrease by similar amounts, at least within CO-detected areas of each complex. The result is a surprisingly stable CO-to-\ipah\ ratio. Meanwhile the drop in CO-to-\idust\ may indicate that the CO molecule (like the PAHs producing \ipah) is more fragile and sensitive to metallicity than the grains that produce the \idust. In that sense, the CO-to-\idust\ ratio may reflect the CO-to-dust ratio, which has often been observed to be lower in low metallicity dwarf galaxies and used as a way to infer the CO-to-H$_2$ conversion factor \citep[e.g.,][]{leroy2011,shi2015,chiang2024}.

The fields in NGC~6822 do not show identical results. Fields 2 and 4 show the highest fraction of MIR emission coincident with CO and also the lowest CO-to-MIR ratios in both bands. These are also the two fields with the most extensive CO detections and the highest \iratio\ ratios. As discussed by \citet{schruba2017} these complexes are in distinct evolutionary stages. Our measurements show Fields 2 and 4 in a phase with more abundant CO emission, abundant PAHs relative to small dust, and spatially concentrated bright MIR emission.

A region in the southern part of Field 4 is an interesting exception to the overall correlation between CO and \ipah. There, we find a region with detected CO and \ipah, and a high CO/F770W ratio of about 20.

\subsection{MIR emission, atomic gas, and total neutral gas} \label{sec:mir_gas}

Finally, we compare F770W and F2100W emission to atomic and total gas column density. High column density \ion{H}{1} is present over the whole area of all of our fields and represents the majority of the total gas along most lines of sight. Based on \citet{draine2007, compiegne2008, draine2011}, to first order we expect
\begin{equation}\label{eq:ipah_htot}
\ipah/N(\mathrm{H_{tot}})\propto \mathrm{DGR}\times U\times \qpah,
\end{equation}
and
\begin{equation}\label{eq:idust_htot}
\idust/N(\mathrm{H_{tot}})\propto \mathrm{DGR}\times U,
\end{equation}
with $U$ the interstellar radiation field intensity, DGR the dust to gas ratio, $\qpah$ the PAH abundance, and $N(\mathrm{H_{tot}})$ the total gas column density 
\begin{equation}\label{eq:nhtot}
    N(\mathrm{H_{tot}}) \equiv N(\mathrm{HI}) + 2N(\mathrm{H_2}).
\end{equation}

\begin{figure*}
\begin{center}
\includegraphics[width=\textwidth]{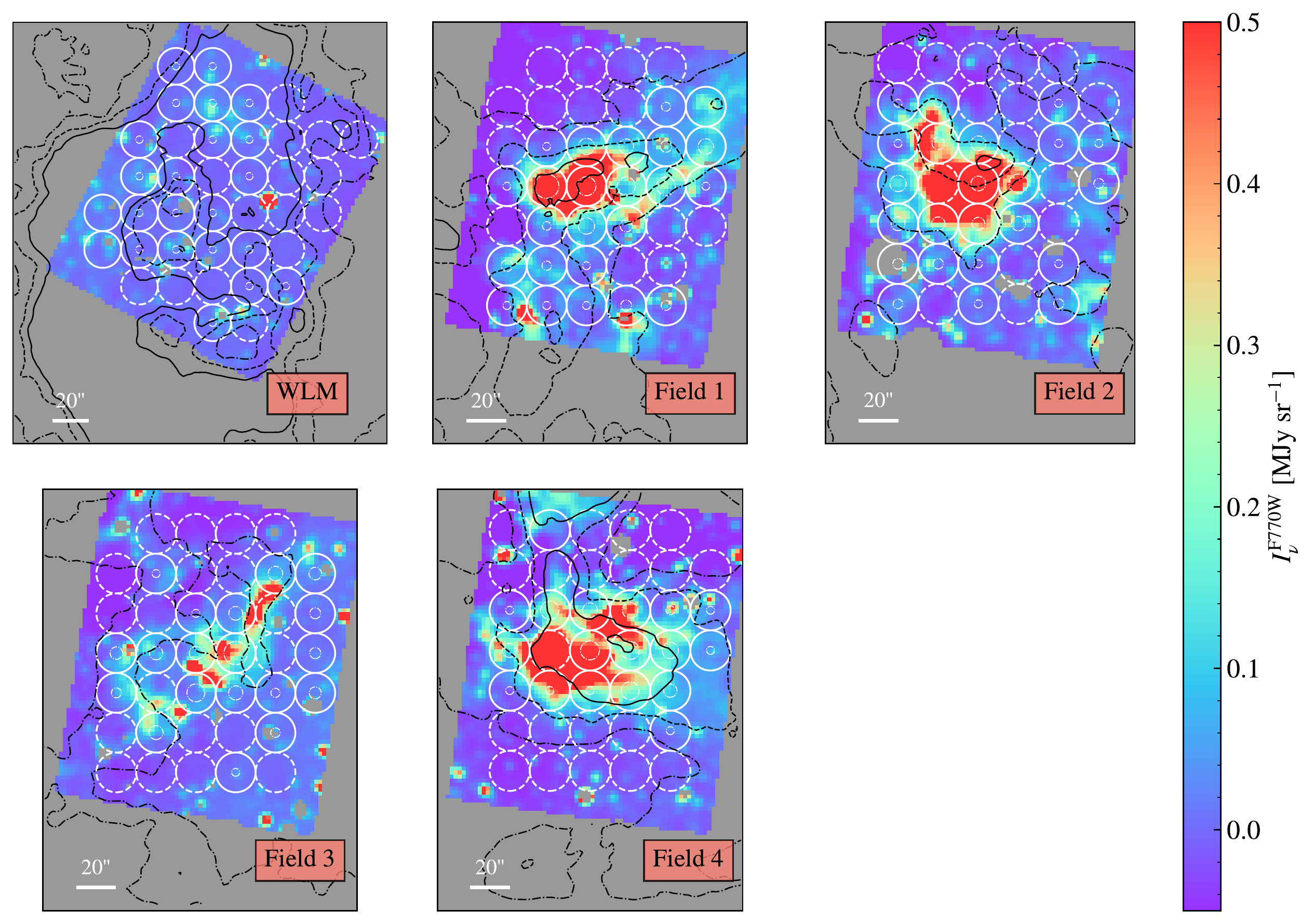}
\caption{JWST F770W (top) and F2100W (bottom) images of WLM and NGC~6822 Fields 1--4 with \hone\ column density contours of $10^{21}$~cm$^{-2}$ (dot-dashed),  $1.7\times10^{21}$~cm$^{-2}$ (dashed), and $2.2\times10^{21}$~cm$^{-2}$ (thick solid). The circular apertures ($10''$ radius) were selected such that they sample variations in mid-IR and \hone\ inside and outside the \hplus\ regions (see \S\ref{sec:mir_gas}). Each $10''$ aperture contains additional information about the $\ipah/N(\mathrm{H_{tot}})$ ratio in that aperture. If the radio is undetected (upper limits in  Fig.~\ref{fig:mir_vs_nh_and_nhtot}), the aperture has a dashed boundary, otherwise it has a solid boundary. Detected apertures have an inner circular aperture indicating the value of the measured $\ipah/N(\mathrm{H_{tot}})$ ratio, with a radius of $10''$ corresponding to $\log_{10}\ipah/N(\mathrm{H_{tot}})=-20.5$, and a radius of $1''$ corresponding to $-23$. One can see that the largest ratios are found in the brightest parts of the complexes. Interestingly, the middle of the bubble of Field 3 shows a non-detection in the ratio, and is suppressed in both gas content and PAH emission. The average properties of these apertures are shown in Fig.~\ref{fig:mir_vs_nh_and_nhtot}. A version of this plot with F2100W instead of F770W is in  Fig.~\ref{fig:f2100w_apertures}.
\label{fig:f770w_apertures}}
\end{center}
\end{figure*}

We place $10''$ radius circular apertures that span each field (Fig.~\ref{fig:f770w_apertures}). This aperture size was chosen to match the FWHM~$\approx8''$ beam of our \hone\ maps. Within each aperture, we measure the median \ipah , median \idust , the \hone\ column density ($N(\mathrm{HI})$), and total H column density (Eq.~\ref{eq:nhtot}), assuming the CO-to-\htwo\ conversion factors in Sect.~\ref{subsec:data_alma}. We use the median when calculating these aperture-averaged intensities in order to suppress the contribution of MIR point sources (\S \ref{subsec:ptsrc}).

\begin{figure*}
\begin{center}
\includegraphics[width=\textwidth]{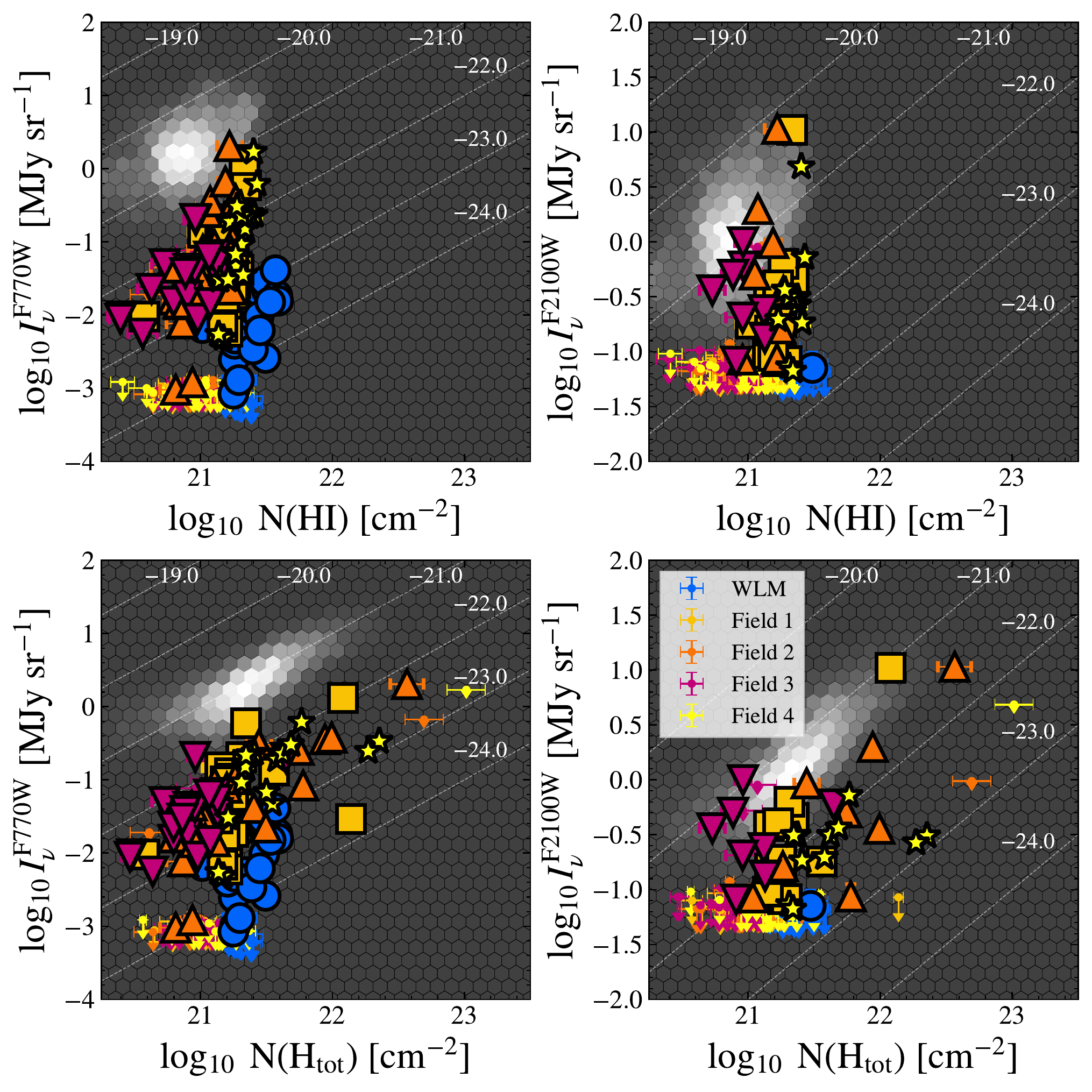}
\caption{Relationships between average mid-IR emission and \hone\ column density (top row), and total H column density (bottom row) assuming appropriate $\alpha_\mathrm{CO}$ (Section~\ref{subsec:data_alma}). Vertical error bars are the measurement uncertainty, while horizontal error bars indicate scatter. Data points are for apertures on the images, selected by eye to capture variations in relative MIR, \hone, and CO surface brightnesses (\S\ref{sec:mir_gas}). Diagonal lines show constant values of $\log_{10}(I_\nu^\mathrm{MIR}/N(\mathrm{HI}))$ (top) or $\log_{10}(I_\nu^\mathrm{MIR}/N(\mathrm{H_{tot}}))$ (bottom) ratios (in units of $\log_{10}~\mathrm{(MJy~sr^{-1})/cm^{-2}}$).
Detections are indicated as filled symbols with black borders, while $3\sigma$ upper limits of non-detections are smaller circles with arrows. The two-dimensional background histogram shows number counts of \hplus\ regions in PHANGS galaxies; for these regions, a  metallicity-dependent $\alpha_\mathrm{CO}$ was used in the bottom row. At fixed gas column density, PAH emission (left column) in NGC~6822 and WLM is significantly weaker than for \hplus\ regions in PHANGS galaxies. Meanwhile the \idust\ vs. N(\hone) and \idust\ vs. N(H$_\mathrm{tot}$) relationships may be much more similar between PHANGS and NGC~6822 and WLM, but deeper F2100W observations are required to confirm this.
\label{fig:mir_vs_nh_and_nhtot}}
\end{center}
\end{figure*}

The second and third rows of Fig.~\ref{fig:co_mir_ratios} show the MIR-to-$N(\mathrm{HI})$ and MIR-to-$N(\mathrm{H_{tot}})$ ratios for each aperture and Fig.~\ref{fig:mir_vs_nh_and_nhtot} compares MIR intensity and gas column via scatter plots. In NGC~6822, $\ipah$-detected apertures show $\ipah/N(\mathrm{H_{tot}})$ well below those in PHANGS galaxies. Apertures with detected MIR emission in WLM show even lower $\ipah/N(\mathrm{H_{tot}})$ ratios than NGC~6822. The right panels show a similar situation for $\idust / N(\mathrm{H_{tot}})$, with the median ratios of detected apertures lower in NGC6822 compared to PHANGS and again even lower in WLM than in NGC~6822. Each of our fields also shows a significant number of apertures with only an upper limit for \ipah , and these limits imply even lower $\ipah / N(\mathrm{H_{tot}})$ than we measure for detected apertures. Upper limits are even more common for \idust\ than \ipah , reflecting the higher noise in the F2100W data so that this ratio shows even more upper limits. These measurements match the visual impression that atomic gas covers the whole field, while extended PAH and dust emission are fainter and patchier.

Fig.~\ref{fig:mir_vs_nh_and_nhtot} shows the same measurements via scatter plots of \ipah\ and \idust\ as a function of $N(\mathrm{HI})$ and $N(\mathrm{H_{tot}})$. Compared to PHANGS measurements for \ion{H}{2} regions (grayscale), both of our targets show similar high column densities, but significantly lower \ipah\ and moderately lower \idust\ at fixed column density compared to PHANGS. Despite higher column densities WLM shows even lower intensity in both bands compared to NGC~6822. The figure again highlights the large number of apertures with upper limits. It also shows how the apertures with relatively high $\ipah / N(\mathrm{H_{tot}})$ or $\idust / N(\mathrm{H_{tot}})$ are also those with high $N(\mathrm{H_{tot}})$. These tend to be the peaks of the complexes, which also harbor the detected CO emission. 

The measurements in Fig.~\ref{fig:mir_vs_nh_and_nhtot} are shown in Table~\ref{tab:gasdust1}, and again in Table~\ref{tab:gasdust2} but normalized to the PHANGS values, next to estimates of the fractional change in the ratios expected given changes in DGR and \qpah\ (Eq.~\ref{eq:ipah_htot} and Eq.~\ref{eq:idust_htot}). The colored dashed lines in the bottom panels of Fig.~\ref{fig:co_mir_ratios} show the fiducial expectations in Table~\ref{tab:gasdust2}. 

\begin{deluxetable*}{lllllll}
\tablecolumns{7}
 \tablecaption{CO(2-1)-to-MIR ratios, MIR-to-\hone, and MIR-to-$N(\mathrm{H_{tot}})$ in NGC~6822 and WLM, compared to PHANGS.}
    \tablehead{
         \colhead{Target(s)} & \colhead{CO(2-1)/F770W} & \colhead{CO(2-1)/F2100W} & \colhead{F770W/$N(\mathrm{HI})$} & \colhead{F2100W/$N(\mathrm{HI})$} & \colhead{F770W/$N(\mathrm{H_{tot}})$} & \colhead{F2100W/$N(\mathrm{H_{tot}})$}\\
         \colhead{(1)} & \colhead{(2)} & \colhead{(3)} & \colhead{(4)} & \colhead{(5)} & \colhead{(6)} & \colhead{(7)}}
\startdata
PHANGS & $-0.04^{+0.66}_{-0.20}$ & $0.03^{+0.61}_{-0.23}$ & $-21.07^{+0.54}_{-0.56}$ & $-21.13^{+0.51}_{-0.51}$ & $-21.24^{+0.40}_{-0.49}$ & $-21.35^{+0.42}_{-0.45}$ \\
IC 5332 \& NGC5068 & $-0.23^{+1.20}_{-0.15}$ & $-0.16^{+1.18}_{-0.20}$ & $-21.37^{+0.56}_{-0.59}$ & $-21.33^{+0.47}_{-0.56}$ & $-21.52^{+0.57}_{-0.67}$ & $-21.60^{+0.55}_{-0.57}$ \\
NGC~6822 & $0.17^{+0.01}_{-0.01}$ & $-0.68^{+0.01}_{-0.01}$ & $-22.35^{+0.35}_{-0.35}$ & $-21.75^{+0.30}_{-0.30}$ & $-22.46^{+0.27}_{-0.27}$ & $-21.98^{+0.29}_{-0.29}$ \\
... Field 1  & $0.22^{+0.01}_{-0.01}$ & $-0.57^{+0.01}_{-0.01}$ & $-22.45^{+0.48}_{-0.50}$ & $-21.98^{+0.40}_{-0.25}$ & $-22.52^{+0.44}_{-0.52}$ & $-21.98^{+0.41}_{-0.32}$ \\
... Field 2  & $-0.113^{+0.004}_{-0.004}$ & $-1.109^{+0.004}_{-0.004}$ & $-22.56^{+0.89}_{-0.26}$ & $-21.53^{+0.70}_{-0.50}$ & $-22.58^{+0.31}_{-0.38}$ & $-22.05^{+0.50}_{-0.35}$ \\
... Field 3  & $0.39^{+0.04}_{-0.04}$ & $-0.39^{+0.04}_{-0.04}$ & $-22.37^{+0.17}_{-0.32}$ & $-21.65^{+0.47}_{-0.35}$ & $-22.40^{+0.13}_{-0.27}$ & $-21.74^{+0.57}_{-0.26}$ \\
... Field 4  & $0.184^{+0.006}_{-0.006}$ & $-0.652^{+0.006}_{-0.006}$ & $-22.04^{+0.25}_{-0.69}$ & $-21.86^{+0.18}_{-0.35}$ & $-22.35^{+0.31}_{-0.53}$ & $-22.14^{+0.25}_{-0.50}$ \\
WLM  & $-0.48^{+0.04}_{-0.04}$ & $-1.12^{+0.05}_{-0.05}$ & $-23.46^{+0.38}_{-0.51}$ & $-22.49^{+0.15}_{-0.15}$ & $-23.58^{+0.53}_{-0.45}$ & $-22.48^{+0.14}_{-0.14}$ \\
\enddata
\tablecomments{All quantities are expressed in $\log_{10}$ units, and uncertainties are 16th-84th percentiles. Columns 2 and 3 are $\log_{10}I_\mathrm{CO(2-1)}/I_\nu^\mathrm{MIR}$ ratios in units of K~km~s$^{-1}$/(MJy~sr$^{-1}$), as defined in Eq.~\ref{eq:r_comir}. Columns 4--7 are $\log_{10}I_\nu^\mathrm{MIR}/N(\mathrm{H_{tot}})$ ratios in units of MJy~sr$^{-1}$/cm$^{-2}$. Columns 2 and 3 are plotted in the first row of Fig.~\ref{fig:co_mir_ratios}, while columns 4--7 are shown in the second and third rows of that figure. See \S\ref{sec:mir_co} and \S\ref{sec:mir_gas} for details.
\tablenotetext{a}{The entry for CO/F770W (column 2) can be compared with the value of $\log_{10}$CO(2-1)/F770W (not starlight subtracted) from Table 2, row 3 of \citet{chown2025}, which is $0.00\pm0.35$~dex. The (statistically insignificant) difference in the median ratio is because our analysis is restricted to the 19 PHANGS Cycle 1 Treasury targets while that analysis used 70 targets (Cycles 1 + 2). 
}
}
\label{tab:gasdust1}
\end{deluxetable*}

\begin{deluxetable*}{lllll}
\tablecolumns{5}
\tablecaption{$\log_{10}I_\nu^\mathrm{MIR}/N(\mathrm{H_{tot}})$ ratios for F770W and F2100W, normalized by their respective ratios measured from PHANGS galaxies (row 1 of Table~\ref{tab:gasdust1}), with dust-to-gas ratio and PAH abundances normalized to PHANGS values for comparison.}
\tablehead{
     \colhead{Target(s)} & \colhead{$\Delta\log_{10}\frac{\ipah}{N(\mathrm{H_{tot}})}$} & \colhead{$\Delta\log_{10} \mathrm{DGR}~\qpah$} & \colhead{$\Delta\log_{10}\frac{\idust}{N(\mathrm{H_{tot}})}$} & \colhead{$\Delta\log_{10} \mathrm{DGR}$} \\
     \colhead{(1)} & \colhead{(2)} & \colhead{(3)} & \colhead{(4)} & \colhead{(5)}}
\startdata
NGC~6822 & $-1.22^{+0.43}_{-0.53}$ & $-1.04$ & $-0.63^{+0.47}_{-0.49}$ & $-0.49$ \\
... Field 1  & $-1.27^{+0.60}_{-0.71}$ & \ldots & $-0.63^{+0.59}_{-0.56}$ & \ldots \\
... Field 2  & $-1.34^{+0.51}_{-0.62}$ & \ldots & $-0.70^{+0.65}_{-0.58}$ & \ldots \\
... Field 3  & $-1.16^{+0.42}_{-0.55}$ & \ldots & $-0.39^{+0.70}_{-0.52}$ & \ldots \\
... Field 4  & $-1.10^{+0.51}_{-0.72}$ & \ldots & $-0.79^{+0.49}_{-0.68}$ & \ldots \\
WLM  & $-2.34^{+0.67}_{-0.66}$ & $-1.64$ & $-1.13^{+0.44}_{-0.47}$ & $-0.79$ \\
\enddata
\tablecomments{Columns 2 and 4 show columns 5 and 7 of Table~\ref{tab:gasdust1} divided by the first row of that table. The values here are the large diamonds in the bottom row of Fig.~\ref{fig:co_mir_ratios} divided by the value of the black dashed line. The first row is the mean of Fields 1--4 in NGC~6822. Guided by Equations~\ref{eq:ipah_htot} and \ref{eq:idust_htot}, column 3 (to be compared with column 2) shows the product of dust-to-gas ratio (DGR) and PAH abundance \qpah\ divided by their PHANGS values (see note in Table~\ref{tab:gal} for how DGR was estimated), and column 5 (to be compared with column 4) shows the same but with just DGR. Since we do not constrain DGR or \qpah\ here, we use galaxy-integrated values. This is illustrated in the bottom row of Fig.~\ref{fig:co_mir_ratios}. See \S\ref{sec:mir_gas} for details.}
\label{tab:gasdust2}
\end{deluxetable*}

In NGC~6822, the median $\log_{10}\ipah/N(\mathrm{H_{tot}})$ and $\log_{10}\idust/N(\mathrm{H_{tot}})$ for detected apertures appear overall consistent with the fiducial expectation. WLM shows lower ratios, consistent with the true DGR or \qpah\ in the diffuse ISM of WLM being even lower than our adopted values. As we remark above, the WLM MIR images appear largely devoid of extended emission even though these observations are deeper than those for NGC~6822 and the \ion{H}{1} column densities are higher in this target. WLM's high inclination may also play a role here, as high inclination leads to high \ion{H}{1} column densities, but will also lead to the inclusion of gas from the extended \ion{H}{1} reservoir in each measurement, including the outskirts of the galaxy.

Both fields show many ($3\sigma$) upper limits. For F770W, the majority of the upper limits indicate  $\log_{10}\ipah/N(\mathrm{H_{tot}})$ significantly below the measured detections. As noted in \S\ref{subsec:data_jwst}, our tests show that the background  level of the JWST observations is uncertain at the $\approx 0.1$~MJy~sr$^{-1}$ level. Our quoted upper limits which are derived from estimates of the instrumental noise, do not account for this term. Fig. \ref{fig:mir_vs_nh_and_nhtot} shows that even in the extreme case where we considered all intensities below $0.1$~MJy~sr$^{-1}$ ($\log_{10} I_\nu = 0.1$) as upper limits, both targets would lie well below the PHANGS targets in both bands.

\begin{figure*}
\begin{center}
\includegraphics[width=\textwidth]{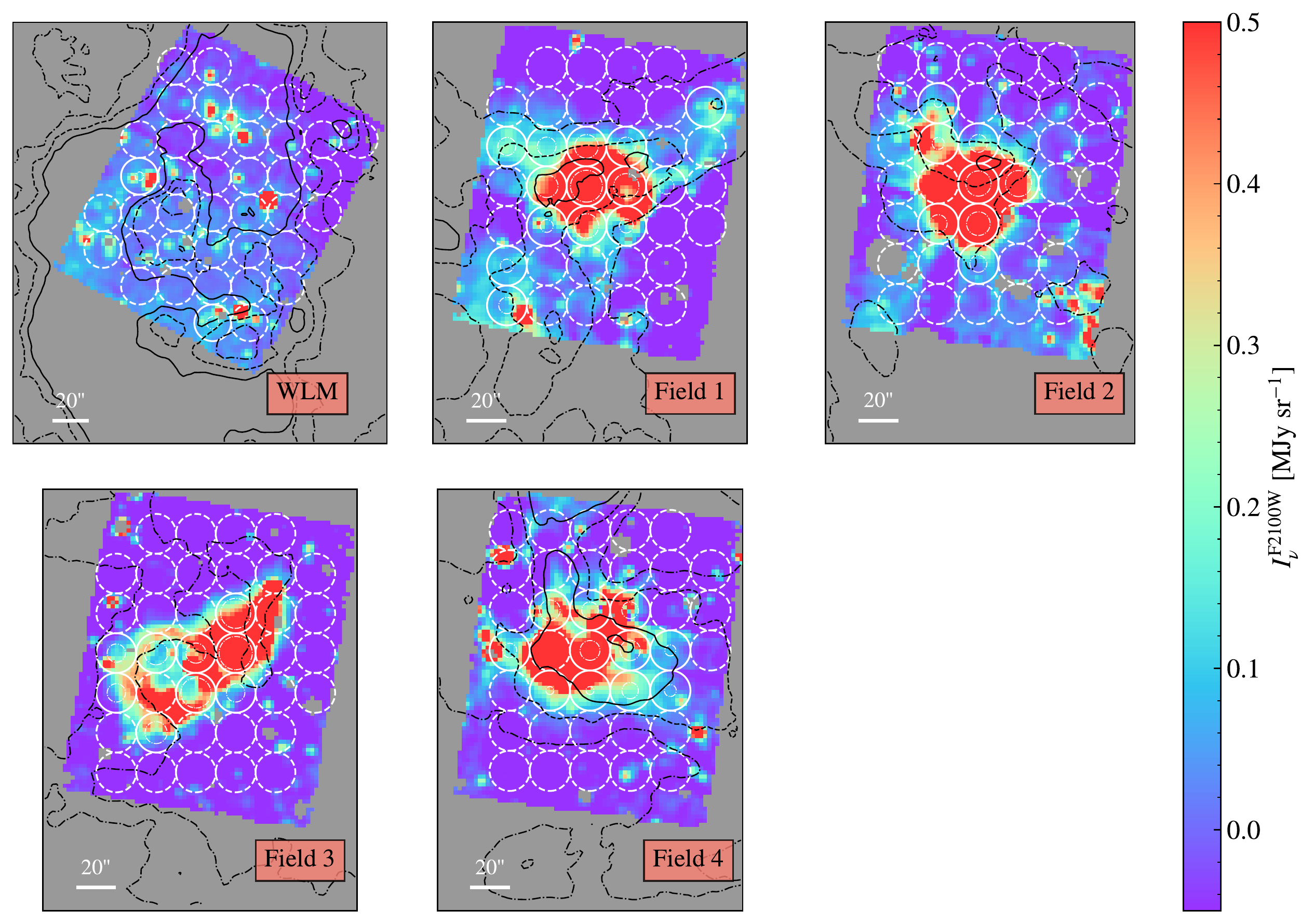}
\caption{As in Fig.~\ref{fig:f770w_apertures} but with F2100W instead of F770W. 
\label{fig:f2100w_apertures}}
\end{center}
\end{figure*}

Figure~\ref{fig:f770w_apertures} shows  $\ipah/N(\mathrm{H_{tot}})$ for individual apertures. Detections are shown as solid circles, while non-detections have dashed boundaries and no inner apertures. Inside each detected aperture, the size of the second, inner circle is proportional to $\log_{10}\ipah/N(\mathrm{H_{tot}})$. The largest ratios follow the bright MIR emission, while the outskirts of each field show many non-detections. The figure shows that fainter, extended F770W emission follows the distribution of \hone\ emission in each complex. The center of the bubble in Field 3 provides a striking example of this correlation, showing depressed \hone\ emission, faint or undetected F770W, and lower $\ipah/N(\mathrm{H_{tot}})$ ratio than the apertures around the border of the bubble. In addition to the destruction of PAHs within the ionized gas in the center of the bubble, this shows pile-ups of both gas and dust at the edges of the expanding bubble. The version of this plot with F2100W in place of F770W is shown in  Fig.~\ref{fig:f2100w_apertures}.

Finally, we note that $U$, the intensity of the interstellar radiation field appears in Eqs. \ref{eq:ipah_htot} and \ref{eq:idust_htot}. All other things being equal, MIR emission should be brighter in regions with more intense radiation. We currently lack resolved estimates of $U$ within our regions, but note that it should be high near the massive O and B stars powering the regions. We note that the diffuse interstellar radiation field in the SMC appears consistent with or higher than the intensity of the Solar Neighborhood field \citep[][]{utomo2019,chastenet2019}, so that there is no evidence that dust or PAH emission will ``hide'' due to low $U$ fields in these environments.

\section{Conclusions}\label{sec:conclusions}

With the goal of constraining the evolution of small dust grains at low metallicity and exploring the utility of PAH and small dust grain emission to trace cold gas, we have used JWST/MIRI to image mid-infrared (MIR) emission from two of the closest star-forming, low-metallicity dwarf galaxies, NGC~6822 ($Z=0.2~Z_\odot$, $D=474$~kpc) and the Wolf–Lundmark–Melotte galaxy (WLM, $Z=0.13~Z_\odot$, $D=985$~kpc). We present new 0.7--3.3 pc resolution F770W  and F2100W images that cover the main star-forming regions in each galaxy (four regions in NGC~6822 and one in WLM). In more massive galaxies, F770W observations are dominated by the 7.7~\um\ PAH band, while the F2100W filter is dominated by emission from stochastically-heated small dust grains at $\approx 21$~\um.

The MIR images of NGC~6822 (Figures~\ref{fig:rgb_6822_1} and \ref{fig:rgb_6822_3}) reveal bright emission with detailed structure. We observe filaments, edge-brightened bubbles, extended diffuse emission, and a plethora of point sources of various types. The MIR emission in NGC~6822 appears bright with extended structure, but is still confined relative to the diffuse atomic gas \citep[visible from new LGLBS VLA observations][E. Koch et al. submitted]{pingel2024}. This atomic gas extends at high column densities across each field and far beyond (Fig.~\ref{fig:overview}), while the MIR emission appears mostly confined to the fields covered by our observations. In contrast to NGC~6822, MIR emission from the lower metallicity galaxy WLM (Fig.~\ref{fig:rgb_wlm}) shows barely any extended emission, with dust emission mostly limited to a smaller number of point sources. 

We compare these images to one another, to maps of CO and 21-cm emission tracing the cold gas, to \textit{Spitzer} observations of the Magellanic Clouds, and to PHANGS-JWST observations of more massive galaxies, and find:

\begin{enumerate}
   
\item Employing \iratio\ as an empirical tracer of the PAH abundance (see Appendix~\ref{sec:appendix_qpah}), we find that NGC~6822 and WLM both show suppressed PAH abundance compared to more massive, nearly solar metallicity galaxies in PHANGS-JWST (Figures~\ref{fig:violins}, \ref{fig:7_vs_21}, Tables \ref{tab:stats}, \ref{tab:fits}). We find substantial, $\approx 0.5$~dex, field-to-field variation in NGC~6822 with high values in Field 4 similar to the SMC, and low values in Fields 1--3 similar to WLM. 
These measurements
appear consistent with the well-documented drop in PAH emission at low metallicity found in numerous \textit{Spitzer} studies.

\item Our observations resolve the \hplus\ regions within the larger star-forming complexes. In all four NGC~6822 fields, we observe decreased \iratio\ within the \hplus\ regions compared to the surrounding clouds, consistent with PAH destruction in ionized gas. In NGC~6822, we observe a drop of $0.50$~dex in the \iratio\ ratio from outside to inside the \hplus\ regions (Table~\ref{tab:stats}). This is consistent with the drop in the \textit{Spitzer} $I_\nu^\mathrm{IRAC~8~\mu m}/I_\nu^\mathrm{MIPS~24~\mu m}$ ratio or fitted PAH abundance seen between the neutral and ionized gas the Magellanic Clouds and in the drop in PAH abundance measured contrasting \hplus\ regions and diffuse ISM at larger scales in PHANGS-JWST (Fig.~\ref{fig:7_vs_21}, Table~\ref{tab:stats}, Appendix~\ref{sec:appendix_qpah}).

\item Because \hplus\ regions still exhibit bright MIR emission, our measured drop in \iratio\ towards their interiors leads to broken power-law when plotting \ipah\ vs. \idust\ intensity, with a shallow slope ($<1$) at high intensities. This relationship appears common among NGC~6822, the Magellanic Clouds, and the PHANGS-JWST galaxies.

\item Within the well-shielded, few-pc-sized regions where CO emission is detected in both NGC~6822 and WLM, we observe $I_\mathrm{CO(2-1)}$-to-\ipah\ ratios similar to those observed at $50{-}100$~pc scales in more massive galaxies at higher metallicity. Meanwhile, the $I_\mathrm{CO(2-1)}$-to-\idust\ ratio even in these CO-bright regions appears significantly lower in NGC~6822 and WLM compared to more massive galaxies. This suggests that CO and PAH emission may be suppressed by comparable amounts at these metallicities, but that the small, stochastically heated dust grains traced by the F2100W filter remain more abundant than CO and PAHs. We emphasize that on average, only $5$--$50$\% (F770W) and $6$--$74$\% (F2100W) of the MIR emission appears coincident with CO detections and enters this analysis (Table~\ref{tab:ptsrc}). While the ALMA data lack short spacing and sensitivity to compare rigorously to JWST over a wider area, qualitatively both the F770W and F2100W appear more extended than the ALMA CO.

\item While all of our target fields are full of high column density $21$-cm \hone\ emission, the MIR emission appears more spatially confined, dramatically more so in WLM. Compared to PHANGS galaxies, we observe low \idust\-to-\hone\ column density ratios and even lower \ipah-to-\hone\ ratios, reflective of low abundance of small dust grains in the diffuse ISM of NGC~6822 and WLM. In detected apertures in NGC 6822, these ratios appear consistent with being suppressed by factors proportional to the reduced dust-to-gas ratios and PAH abundances in that galaxy, though there are also many apertures where we measure only an upper limit. WLM shows evidence for additional suppression of dust emission relative to these expectations.

\end{enumerate}

These results paint a dynamic picture of the life cycle of small dust grains, PAHs, and gas in galaxies across metallicity. At the same time, these data leave much to be explored in subsequent work, some of which requires new observations, including detailed characterization of point sources, 
more detailed investigation of the PAH properties (e.g., size and charge), and accounting for the effects of a varying interstellar radiation field within our regions. 

\begin{acknowledgments}

R.~I. acknowledges support from JWST-GO-02130.006-A. V.~V. acknowledges support from the ANID BASAL project FB210003. S.C.O.G. acknowledges financial support from the European Research Council via the ERC Synergy Grant ``ECOGAL'' (project ID 855130) and from the German Excellence Strategy via the Heidelberg Cluster of Excellence (EXC 2181 - 390900948) ``STRUCTURES''.
JC acknowledges funding from the Belgian Science Policy Office (BELSPO) through the PRODEX project ``JWST/MIRI Science exploitation'' (C4000142239).

This paper makes use of the following ALMA data: ADS/JAO.ALMA\#2018.1.00337.S, ADS/JAO.ALMA\#2012.1.00336.S, and ADS/JAO.ALMA\#2024.1.01179.S. ALMA is a partnership of ESO (representing its member states), NSF (USA) and NINS (Japan), together with NRC (Canada), NSTC and ASIAA (Taiwan), and KASI (Republic of Korea), in cooperation with the Republic of Chile. The Joint ALMA Observatory is operated by ESO, AUI/NRAO and NAOJ.

This work was conducted as part of the Local Group L-Band survey (LGLBS).

The LGLBS is an Extra Large program conducted on Jansky Very Large Array, which is operated by the National Radio Astronomy Observatory (NRAO) and includes observations from VLA projects 20A-346, 13A-213, 14A-235, 14B-088, 14B-212, 15A-175, 17B-162, and GBT projects 09A-017, 13A-420, 13A-430, 13B-169, 14A-367, 16A-413. 

NRAO is a facility of the National Science Foundation operated under cooperative agreement by Associated Universities, Inc. Execution of the LGLBS survey science was supported by NSF Award 2205628.

\end{acknowledgments}

\vspace{5mm}
\facilities{JWST, ALMA, VLA, Spitzer}

\appendix

\section{What \qpah\ is implied by our measured \iratio ?}\label{sec:appendix_qpah}

We measure variations in \iratio , which is commonly interpreted as a tracer of fraction of the dust mass in PAHs. Because there are multiple strong PAH features and the far-IR SED, not the MIR, offers the most direct tracer of large dust grain mass, one would ideally prefer to use the ratio of total PAH-to-total IR luminosity or full IR SED modeling to estimate \qpah\ \citep[][and see \citealt{galliano2018, li2020a}]{draine2007a}. Lacking the  data required for those approaches one can still estimate \qpah\ via \iratio\ given models and assumptions about the interstellar radiation field (ISRF) and dust property distribution.
Following similar work in PHANGS by \citep{chastenet2023, sutter2024}, here we compare our measured \iratio\ to model calculations by \citet{draine2007a} and \citet{hensley2023}, and to results fitting SED models to measurements with lower physical resolution but more complete wavelength coverage by \citet{chastenet2025}.

\begin{figure*}
\begin{center}
\includegraphics[width=\textwidth]{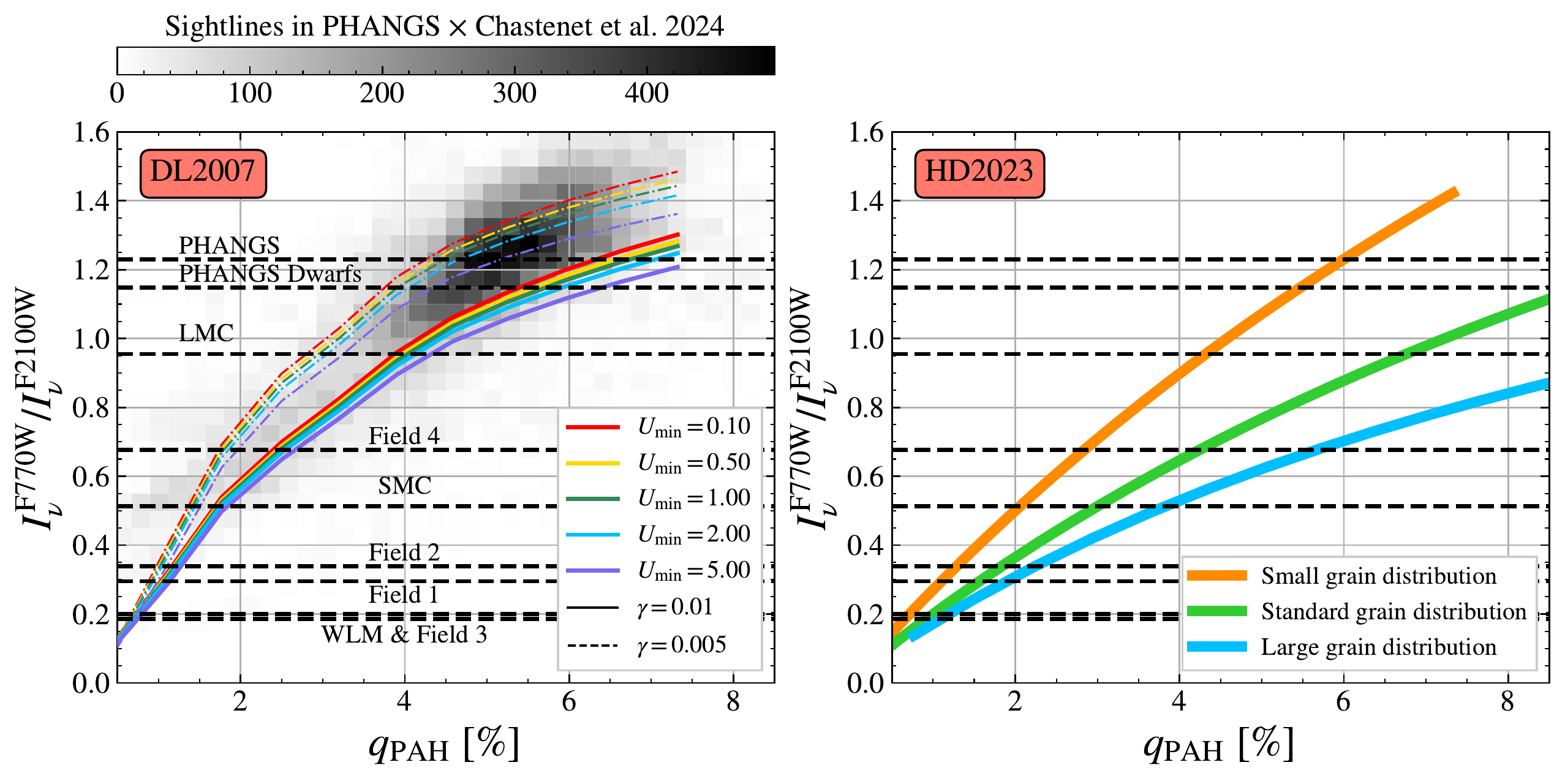}
\caption{\iratio\ as a function PAH-to-dust mass fraction \qpah . The colored curves in the left panel show the \citet{draine2007} dust models with varying ISRF distributions ($U_\mathrm{min}$ is the intensity of the diffuse ISRF, $\gamma$ indicates the fraction of dust heated by more intense radiation). The curves in the right panel were computed using the \citet{hensley2023} dust model which adopts the PAH population from \citet{draine2021}. The three curves represent the ``small", ``standard", and ``large" grain distributions ($a_{01}=3,4,5$~\AA). Grayscale in the left image show data density for $\approx 20''$ regions in $33$ PHANGS targets where \citet{chastenet2025} estimate \qpah\ from IR SED modeling using the \citet{draine2007a} models and we measure \iratio\ at matched resolution. These therefore give a sense of the \iratio{-}\qpah\ relation for real parts of galaxies for these models. The horizontal lines show the median \iratio\ for samples studied in this paper (Table~\ref{tab:stats}; N.B. the  LMC and SMC show the slightly different $I_\nu^\mathrm{IRAC~8\mu m}/I_\nu^\mathrm{MIPS~24\mu m}$).
\label{fig:qpah}}
\end{center}
\end{figure*}

In Figure~\ref{fig:qpah} we show  \qpah\ as a function of \iratio , marking the values of \iratio\ measured in our data. Colored solid lines show model curves from \citet{draine2007} and \citet{hensley2023}. The left panels illustrate how the \iratio{-}\qpah\ relationship depends on aspects of the ISRF. We vary the diffuse ISRF intensity, $U_\mathrm{min}$, and the fraction of dust mass illuminated by more intense radiation fields, $\gamma$ \citep{draine2007a}. The right panel shows the \citet{hensley2023} models and illustrates the effect of varying the grain size distributions \citep[following the size distribution definitions in][]{draine2021}. 

In the left panel, the grayscale shows data density for SED-fitting results where \citet{chastenet2025} fit \qpah\ based on combined \textit{Herschel} and WISE data. From their atlas, we selected $34$ PHANGS galaxies that also have JWST F770W and F2100W measurements, which we used to measure \iratio\ at resolution matched to \citet{chastenet2025}. The grayscale thus gives a practical estimate of how \iratio\ traces model-fit \qpah\ in real local star-forming galaxies \citep[see also][]{sutter2024}.

Fig. \ref{fig:qpah} shows that the \iratio\ for WLM and NGC~6822 Fields 1, 2, and 3 lies below almost all of the \citet{chastenet2025} comparison data. Comparing to the \citet{draine2007} curves or extrapolating the \citet{chastenet2025} trend implies low \qpah, of order $\approx 1 \pm 0.5 \%$ for these four fields and the \citet{draine2007} models. This agrees with previous modeling by \citet{draine2007}, which found \qpah of 0.7--1.1\% for the integrated SED of NGC~6822 \citep{draine2007}. Field 4 shows higher \iratio\ and does overlap some of the \citet{chastenet2025} comparison data. On their system, we would expect this field to have $\qpah \approx 2 \pm 0.5\%$. For comparison, the typical PHANGS ratio that we measure corresponds to $\qpah \approx 5 \pm 1\%$ in the \citet{chastenet2025} comparison sample, in good agreement with \citet{sutter2024}. We emphasize that these values are still dependent on the physical dust properties assumed, as illustrated by variation between the models in the left an right panel.

These model curves and the comparison to \citet{chastenet2025} considers filter-integrated measurements, which work well given some model framework, but a more empirical path forward is to add up the PAH luminosity from each of the major features rather than just the 7.7~\um\ feature (i.e., also including the $3.3\mu$m and $11.3\mu$m complexes). In environments where PAH emission becomes faint (such as NGC~6822 and WLM), it will also be important for future work to account for contaminants in the filters, including starlight and continuum from small dust grains in the F770W filter and possibly some PAH emission in the F2100W filter. The fractional contributions are most likely strong functions of location, as seen in other sources \citep[e.g.][]{chown2024b}.

\bibliography{ryan}{}
\bibliographystyle{aasjournal}

\end{document}